\documentclass[11pt,titlepage,twoside,leqno]{article}

\usepackage{eepic,epic}

\usepackage{latexsym}
\usepackage{amsfonts}
\usepackage{amsmath}
\usepackage{amssymb}

\makeatletter

\newcount\hour  \newcount\minutes  \hour=\time  \divide\hour by 60
\minutes=\hour  \multiply\minutes by -60  \advance\minutes by \time
\def\mmmddyyyy{\ifcase\month\or Jan\or Feb\or Mar\or Apr\or May\or Jun\or Jul\or
  Aug\or Sep\or Oct\or Nov\or Dec\fi \space\number\day, \number\year}
\def\hhmm{\ifnum\hour<10 0\fi\number\hour :%
  \ifnum\minutes<10 0\fi\number\minutes}

\def\section{\@startsection {section}{1}{\z@}
    {3.5ex plus1ex minus.2ex}{2.3ex plus.2ex}{\Large\bf}}
\def\subsection{\@startsection{subsection}{2}{\z@}
    {3.25ex plus1ex minus.2ex}{1.5ex plus.2ex}{\large\bf}}
\def\subsubsection{\@startsection{subsubsection}{3}{\z@}
    {3.25ex plus1ex minus.2ex}{1.5ex plus.2ex}{\normalsize\bf}}
\def\paragraph{\@startsection{paragraph}{4}{\z@}
    {3.25ex plus1ex minus.2ex}{1em}{\normalsize\bf}}
\def\subparagraph{\@startsection{subparagraph}{4}{\parindent}
    {3.25ex plus1ex minus.2ex}{1em}{\normalsize\bf}}
\makeatother

\makeatletter \@beginparpenalty=10000 \makeatother
\def\underl#1 {\leavevmode\let\first=\relax\underli #1 }
\def\underli#1 {\ifx&#1\let\next=\relax\unskip
                \else\let\next=\underli\first\ulinebox{#1}\fi\let\first=\undersp\next}
\def\undersp{\penalty50\ulinebox{\space}\penalty50}
\def\ulinebox#1{\vtop{\hbox{\strut#1}\hrule}}
\def\unice#1 {\underl #1 & }

\def\desclabel#1{\bf #1\hfil}
\def\desc{\list{}{%
\labelwidth=\leftmargin
\advance \labelwidth by -\labelsep
\let \makelabel=\desclabel}}

\makeatletter 

\newlength{\filength}
\newsavebox{\gcbox}
\sbox{\gcbox}{\framebox[\filength]{\rule{0ex}{2ex}}}

\newlength{\leftjustindent}
\newlength{\@leftjustindent}
\def\leftjust{\let\\\@leftjustcr\let\end\@endleftjust
  \addtolength{\@leftjustindent}{\leftjustindent}
  \vcenter\bgroup
  \halign\bgroup
    \hbox to\displaywidth{
      \rule{\@leftjustindent}{0ex}$\displaystyle##$\hfill
      }\crcr
}
\def\endleftjust{\crcr\egroup\egroup\endgroup}
\def\@endleftjust#1{\crcr\egroup\egroup\@checkend{#1}\endgroup}
\def\@leftjustcr{\crcr}

\newcommand{\qedblob}{\mbox{\rule[-1.5pt]{5pt}{10.5pt}}}
\def\literalqed{{\ \nolinebreak\hfill\mbox{\qedblob\quad}}}

\def\qed{\literalqed}

\newtheorem{theorem}{Theorem}[section]
\newtheorem{corollary}[theorem]{Corollary}
\newtheorem{lemma}[theorem]{Lemma}

\newtheorem{definition}[theorem]{Definition}
\newtheorem{claim}[theorem]{Claim}

\newtheorem{proposition}[theorem]{Proposition}
\newtheorem{remark}[theorem]{Remark}

\newcommand{\singlespacing}{\let\CS=
\@currsize\renewcommand{\baselinestretch}{1}\tiny\CS}
\newcommand{\singlespacingplus}{\let\CS=
\@currsize\renewcommand{\baselinestretch}{1.25}\tiny\CS}
\newcommand{\doublespacing}{\let\CS=
\@currsize\renewcommand{\baselinestretch}{1.75}\tiny\CS}
\newcommand{\draftspacing}{\let\CS=
\@currsize\renewcommand{\baselinestretch}{2.0}\tiny\CS}
\newcommand{\foospacing}{\let\CS=
\@currsize\renewcommand{\baselinestretch}{1.05}\tiny\CS}

\makeatother

\singlespacingplus

\makeatletter
\clubpenalty=\@highpenalty
\widowpenalty=\@highpenalty
\makeatother

\makeatletter
\newcommand{\niceonespacing}{\let\CS=\@currsize\renewcommand{\baselinestretch}{1.1}\tiny\CS}\newcommand{\nicetwospacing}{\let\CS=\@currsize\renewcommand{\baselinestretch}{1.2}\tiny\CS}
\newcommand{\nicethreespacing}{\let\CS=\@currsize\renewcommand{\baselinestretch}{1.3}\tiny\CS}
\newcommand{\singlespacingplusplus}{\let\CS=\@currsize\renewcommand{\baselinestretch}{1.35}\tiny\CS}
\newcommand{\nicefourspacing}{\let\CS=\@currsize\renewcommand{\baselinestretch}{1.4}\tiny\CS}
\newcommand{\nicefivespacing}{\let\CS=\@currsize\renewcommand{\baselinestretch}{1.5}\tiny\CS}
\newcommand{\nicesixpacing}{\let\CS=\@currsize\renewcommand{\baselinestretch}{1.6}\tiny\CS}
\makeatother

\makeatletter
\def\@cite#1#2{[#1\if@tempswa , #2\fi]}
\makeatother

\makeatletter
\def\@citex[#1]#2{\if@filesw\immediate\write\@auxout{\string\citation{#2}}\fi
  \def\@citea{}\@cite{\@for\@citeb:=#2\do
    {\@citea\def\@citea{,\linebreak[0]}\@ifundefined
       {b@\@citeb}{{\bf ?}\@warning
       {Citation `\@citeb' on page \thepage \space undefined}}%
\hbox{\csname b@\@citeb\endcsname}}}{#1}}
\makeatother

\makeatletter
\def\foobarpt{\textfont\z@\tenrm 
  \scriptfont\z@\ninrm \scriptscriptfont\z@\sevrm
\textfont\@ne\tenmi \scriptfont\@ne\ninmi \scriptscriptfont\@ne\sevmi
\textfont\tw@\tensy \scriptfont\tw@\ninsy \scriptscriptfont\tw@\sevsy
\textfont\thr@@\tenex \scriptfont\thr@@\tenex \scriptscriptfont\thr@@\tenex
\def\unboldmath{\everymath{}\everydisplay{}\@nomath\unboldmath
          \textfont\@ne\tenmi 
          \textfont\tw@\tensy \textfont\lyfam\tenly
          \@boldfalse}\@boldfalse
\def\boldmath{\@ifundefined{tenmib}{\global\font\tenmib\@mbi\@magscale1\global
        \font\tensyb\@mbsy \@magscale1\global\font
         \tenlyb\@lasyb\@magscale1\relax\@addfontinfo\@xiipt
              {\def\boldmath{\everymath
                {\mit}\everydisplay{\mit}\@prtct\@nomathbold
                \textfont\@ne\tenmib \textfont\tw@\tensyb 
                \textfont\lyfam\tenlyb\@prtct\@boldtrue}}}{}\@xiipt\boldmath}%
\def\prm{\fam\z@\tenrm}%
\def\pit{\fam\itfam\tenit}\textfont\itfam\tenit \scriptfont\itfam\ninit
   \scriptscriptfont\itfam\sevit
\def\psl{\fam\slfam\tensl}\textfont\slfam\tensl 
     \scriptfont\slfam\tensl \scriptscriptfont\slfam\tensl
\def\pbf{\fam\bffam\tenbf}\textfont\bffam\tenbf 
   \scriptfont\bffam\ninbf \scriptscriptfont\bffam\ninbf 
\def\ptt{\fam\ttfam\tentt}\textfont\ttfam\tentt
   \scriptfont\ttfam\nintt \scriptscriptfont\ttfam\nintt 
\def\psf{\fam\sffam\tensf}\textfont\sffam\tensf
    \scriptfont\sffam\tensf \scriptscriptfont\sffam\tensf
\def\psc{\@getfont\psc\scfam\@xiipt{\@mcsc\@magscale1}}%
\def\ly{\fam\lyfam\tenly}\textfont\lyfam\tenly 
   \scriptfont\lyfam\ninly \scriptscriptfont\lyfam\sevly
 \@setstrut \rm}

\makeatother

\newcommand{\fp}{{\rm FP}}
\newcommand{\fe}{{\rm FE}}
\newcommand{\sharpp}{{\rm \#P}}
\newcommand{\sharpe}{{\rm \#E}}
\newcommand{\spanp}{{\rm spanP}}

\newcommand{\sharpsat}{{\rm \#SAT}}

\newcommand{\parityp}{{\rm \oplus P}}
\newcommand{\up}{{\rm UP}}

\newcommand{\spp}{{\rm SPP}}

\newcommand{\e}{{\rm E}}

\renewcommand{\ne}{{\rm NE}}
\newcommand{\ue}{{\rm UE}}

\newcommand{\p}{{\rm P}}

\newcommand{\np}{{\rm NP}}

\newcommand{\pp}{{\rm PP}}
\newcommand{\bpp}{{\rm BPP}}

\newcommand{\ph}{{\rm PH}}

\newcommand\PERM{{\rm perm}}

\def\pair#1{{{\langle\!\!~#1~\!\!\rangle}}}

\newcommand{\sigmastar}{\mbox{$\Sigma^\ast$}}
\newcommand{\nats}{\mathbb{N}}
\newcommand{\integers}{\mathbb{Z}}

\newcommand{\census}{{\mbox{\it{}census}}}

\newcommand\seq{\subseteq}

\newcommand\Lolra{\ \Longleftrightarrow \ }

\newcommand{\equalsdef}{\stackrel{\mbox{\protect\scriptsize\rm df}}{=}}

\newenvironment{block}{\begin{list}{\hbox{}}{\leftmargin 1em
    \itemindent -1em \topsep 0pt \itemsep 0pt \partopsep 0pt}}{\end{list}}

\newcommand{\card}[1]{{\vert\,#1\,\vert}}

\foospacing
\title{%
Tally NP Sets and Easy Census Functions
}

{\singlespacing
\author{
Judy Goldsmith\,\thanks{
\protect\singlespacing 
Supported in part 
by NSF grant CCR-9315354.
} \\
Department of Computer Science \\
University of Kentucky \\
Lexington, KY 40506, USA \\
{\tt goldsmit@cs.engr.uky.edu}
\and
Mitsunori Ogihara\,\thanks{
\protect\singlespacing 
Supported in part 
by NSF CAREER Award CCR-9701911.
} \\
Department of Computer Science \\
University of Rochester \\
Rochester, NY 14627, USA \\
{\tt ogihara@cs.rochester.edu}
\and
J\"{o}rg Rothe\,\thanks{
\protect\singlespacing
Supported in part 
by grants
NSF-INT-9513368/\protect\linebreak[0]DAAD-315-PRO-fo-ab and
NSF-CCR-9322513 and
by a NATO Postdoctoral Science Fellowship
from the Deut\-scher Aka\-de\-mi\-scher Aus\-tausch\-dienst
(``Ge\-mein\-sames Hoch\-schul\-sonder\-pro\-gramm~III
von Bund und L\"andern'').
Work done in part while visiting the University of Kentucky and the 
University of Rochester.
} \\ Institut f\"ur Informatik \\
Friedrich-Schiller-Universit\"at Jena \\
07740 Jena, Germany \\
{\tt rothe@informatik.uni-jena.de}
}
}

\date{March 19, 1998}

\makeatletter
\def\@listI{\leftmargin\leftmargini \parsep 4.5pt plus 1pt minus 1pt\topsep
6pt plus 2pt minus 2pt \itemsep  2pt plus 2pt minus 1pt}

\let\@listi\@listI
\@listi
\makeatother

\begin{document}

\typeout{WARNING:  BADNESS used to suppress reporting!  Beware!!}
\hbadness=3000
\vbadness=10000 

\bibliographystyle{alpha}

\pagestyle{empty}
\setcounter{page}{1}

\sloppy

\pagestyle{empty}
\setcounter{footnote}{0}

{\singlespacing

\maketitle

}

\begin{center}
{\large\bf Abstract}
\end{center}
\begin{quotation}
{\singlespacing
\noindent
We study the question of whether every P set has an easy (i.e.,
polynomial-time computable) census function. We characterize this
question in terms of unlikely collapses of language and function
classes such as $\sharpp_1 \seq \fp$, where $\sharpp_1$ is the class
of functions that count the witnesses for tally NP sets. We prove that
every $\sharpp_{1}^{\ph}$ function can be computed in
$\fp^{\sharpp_{1}^{\sharpp_{1}}}$.  Consequently, every P set has an
easy census function if and only if every set in the polynomial
hierarchy does.  We show that the assumption $\sharpp_1 \seq \fp$
implies $\p = \bpp$ and $\ph \seq {\rm MOD}_{k}\p$ for each $k
\geq 2$, which provides further evidence that not all sets in P have
an easy census function. We also relate a set's property of having an
easy census function to other well-studied properties of sets, such as
rankability and scalability (the closure of the rankable sets under
P-isomorphisms).  Finally, we prove that it is no more likely that
the census function of any set in P can be
approximated (more precisely, can be $n^{\alpha}$-enumerated in
time~$n^{\beta}$ for fixed $\alpha$ and~$\beta$) than that
it can be precisely computed in
polynomial time.
    
}
\end{quotation}


\foospacing
\setcounter{page}{1}
\pagestyle{plain}
\sloppy

\section{Introduction}

Does every P set have an easy (i.e., polynomial-time computable)
census function? Many important properties similar to this one were
studied during the past decades to gain more insight into the nature
of feasible computation. Among the questions that were previously
studied are the question of whether or not every P set has an easy to
compute ranking
function~\cite{gol-sip:j:compression,hem-rud:j:ranking}, whether every
P set is P-isomorphic to some rankable
set~\cite{gol-hom:j:scalability}, whether every sparse set in P is
P-printable~\cite{har-yes:j:computation,all-rub:j:print,rao-rot-wat:j:upward},
whether every infinite set in P has an infinite
P-printable subset~\cite{all-rub:j:print,hem-rot-wec:j:easy}, whether
every P-printable set is P-isomorphic to some tally set in
P~\cite{all-rub:j:print}, and whether every P set admits easy
certificate
schemes~\cite{hem-rot-wec:j:easy,hem-rot-wec:c:easy-one-way-permutations},
to name just a few.  Some of those questions arise in the field of
data compression and are related to Kolmogorov complexity, some are
linked to the question of whether one-way functions exist. 

Extending
this line of research, the present paper studies the complexity of
computing the census functions of sets in~P\@. Census functions have
proven to be a particularly important and useful notion in complexity
theory, and their use has had a profound impact upon almost every area
of the field. In particular, this regards
the extensive literature related to the isomorphism conjecture
of Berman and Hartmanis (e.g., \cite{ber-har:j:iso,mah:j:sparse-complete},
and many other papers), 
the work on the existence of Turing-hard
sparse sets (or of polynomial-size circuits) for various complexity classes
(e.g., \cite{kar-lip:c:nonuniform,ko-sch:j:circuit-low,bal-boo-sch:j:sparse,hem-rot:j:boolean}),
the results relating the computation times for NP sets to their 
densities and the results on P-printability~\cite{har-yes:j:computation,all-rub:j:print,rao-rot-wat:j:upward,gol-hom:j:scalability},
the upward separation technique (e.g., 
\cite{har:j:upward,har-imm-sew:j:sparse,all:j:lim,rao-rot-wat:j:upward,hem-jha:j:defying}, see~\cite{hem-hem-hem:jtoappear:downward} for more recent 
advances that are not based on census functions),
the results on positive relativization and relativization to sparse 
oracles (e.g., 
\cite{lon:j:rest,lon-sel:j:sparse,bal-boo-sch:j:sparse}),
the unexpected collapse of the strong exponential-time 
hierarchy~\cite{hem:j:sky},
and applications to extended lowness~\cite{hem-jia-rot-wat:j:join}.

Valiant, in his seminal
papers~\cite{val:j:permanent,val:j:enumeration}, introduced $\sharpp$,
the class of functions that count the solutions of NP problems, and
its tally version $\sharpp_{1}$ for which the inputs are given in
unary. Although $\sharpp_{1}$ has not become as prominent
as~$\sharpp$, it contains a number of quite interesting and important
problems such as the problem {\tt Self-Avoiding Walk}
(see~\cite{wel:b:knots}): Given an integer $n$ in unary, compute the
number of self-avoiding walks on the square lattice having length~$n$
and rooted at the origin. {\tt Self-Avoiding Walk} is a well-known
classical problem of statistical physics and polymer chemistry, and it
is an intriguing open question whether {\tt Self-Avoiding Walk} is
$\sharpp_{1}$-complete (see~\cite{wel:b:knots}).
Known problems complete for $\sharpp_1$~\cite{val:j:enumeration} 
have the form: Given an integer $n$ in unary, compute the number of
graphs having $n$ vertices and satisfying a fixed graph property~$\pi$.

In Section~\ref{sec:characterizations}, 
we will characterize the question of whether every P
set has an easy census function in
terms of collapses of language and function classes that are
considered to be unlikely. In particular, every P set has an easy
census function if and only if $\sharpp_1 \seq \fp$.
The main
technical contribution in Section~\ref{sec:characterizations} 
is Theorem~\ref{thm:stack}:
$\sharpp_{1}^{\ph}$ is contained in $\fp^{\sharpp_{1}^{\sharpp_{1}}}$.
An immediate consequence of this result are upward collapse results of
the form: the collapse $\#_{1}\cdot \p \seq \fp$ implies the collapse
$\#_{1}\cdot \ph \seq \fp$. Thus, every P set has an easy census
function if and only if every set in the polynomial hierarchy 
has an easy census function.
Note that the corresponding upward collapse for the $\#$ operator
applied to the levels of PH follows immediately from the upward
collapse property of the polynomial hierarchy itself: $\# \cdot \p
\seq \fp$ implies $\np = \p$ and thus $\ph = \p$; so, $\# \cdot \ph =
\# \cdot \p \seq \fp$. However, for the $\#_{1}$ operator this is not
so clear, since the assumption $\#_{1}\cdot \p \seq \fp$ merely
implies that all {\em tally\/} NP sets are in~P (equivalently, $\ne =
\e$), from which one cannot immediately conclude that $\#_{1}\cdot
\np$ or even $\#_{1}\cdot \ph$ is contained
in~FP\@.
In fact, Hartmanis, Immerman, and 
Sewelson~\cite{har-imm-sew:j:sparse} show that in some relativized world,
$\ne = \e$ and yet the (weak) exponential-time hierarchy does not
collapse. In light of this result, it is quite possible that the
assumption of all tally NP sets being in P does not force all tally
sets from higher levels of the polynomial hierarchy into~$\p$.

We also show that the assumption $\sharpp_1 \seq \fp$ implies both $\p =
\bpp$ and $\ph \seq {\rm MOD}_{k}\p$ for each $k \geq 2$
(Theorem~\ref{thm:impl}), which provides further evidence that not all
sets in P have a census function computable in polynomial time. We
also relate a set's property of having an easy census function to
other well-studied properties of sets, such as
rankability~\cite{gol-sip:j:compression} and
scalability~\cite{gol-hom:j:scalability}. In particular, though each
rankable set has an easy census function, we show that (even
when restricted to the sets in~P) the converse is not true unless $\p
= \pp$. This expands the result of Hemaspaandra and Rudich that every
P set is rankable if and only if $\p = \pp$~\cite{hem-rud:j:ranking}
by showing that $\p = \pp$ is already implied by the apparently weaker
hypothesis that every P set {\em with an easy census function\/} is
rankable. 

Cai and Hemaspaandra~\cite{cai-hem:j:enum} introduced the notion of
enumerative counting as a way of approximating the value of a
$\sharpp$ function deterministically in polynomial time.
Hemaspaandra and Rudich~\cite{hem-rud:j:ranking} show that every P set
is $k$-enumeratively rankable for some fixed~$k$ 
in polynomial time if and only if $\sharpp = \fp$. They
conclude that it is no more likely 
that one can enumeratively rank all sets in P
than that one can exactly compute their ranking functions in
polynomial time. In Section~\ref{sec:enumeration}, 
we similarly characterize the question of whether the
census function of all P sets is $n^{\alpha}$-enumerable in time
$n^{\beta}$ for fixed constants $\alpha$ and~$\beta$, or equivalently,
whether every $\sharpp_1$ function is $n^{\alpha}$-enumerable
in time~$n^{\beta}$.  We show that this hypothesis 
implies $\sharpp_{1} \subseteq
\fp$, and we thus conclude that it is no more likely that one can
$n^{\alpha}$-enumerate the census function of every P set in
time~$n^{\beta}$ than that one can precisely compute its census function
in polynomial time.  

Finally, Section~\ref{sec:oracles} provides a number of relativization
results. 

\section{Notation and Definitions}

Fix the alphabet $\Sigma = \{0,1\}$. $\sigmastar$ denotes the set of
all strings over~$\Sigma$, and $\Sigma^{+} = \sigmastar \setminus
\{\epsilon\}$, where $\epsilon$ denotes the empty string. For any
string $x \in \sigmastar$, we denote the length of $x$ by~$|x|$. For
any set $L \seq \sigmastar$, the number of strings in $L$ is
denoted~$\card{L}$, and the complement of $L$ in $\sigmastar$ is
denoted~$\overline{L}$. Let $L^{=n}$ (respectively, $L^{\leq n}$)
denote the set of strings in $L$ of length $n$ (respectively, of
length at most~$n$).
As a shorthand, we use $\Sigma^{n}$ to denote
$(\sigmastar)^{=n}$. For any set $L$, the {\em census function
of~$L$}, $\census_{L} : \sigmastar \rightarrow\, \nats$, is defined by
$\census_{L}(1^n) \equalsdef \card{L^{=n}}$,\footnote{\protect\singlespacing
The census function of $L$ at $n$ is often defined as the number of
elements in $L$ of length up to $n$ in the literature.  This definition
and our definition are compatible as long as our computability admits
subtraction.
We also note that we let $\census_{L}$ map strings $1^n$
(as opposed to numbers $n$ in binary notation) to $\card{L^{=n}}$ to
emphasize that the input to the transducer computing $\census_{L}$ is
given in unary.  
} 
and $\chi_{L}$ denotes the {\em characteristic function of~$L$}, i.e.,
$\chi_{L}(x) = 1$ if $x \in L$, and $\chi_{L}(x) = 0$ if $x \not\in
L$. 
A set $S$ is said to be {\em sparse\/} if there is a polynomial $p$
such that for each length~$n$, $\census_{S}(1^n) \leq p(n)$.
A set $T$ is said to be {\em tally\/} if $T \seq \{1\}^\ast$.
To encode pairs of strings, we use a one-one, onto pairing
function, $\pair{\cdot , \cdot} : \sigmastar \times \sigmastar
\rightarrow\, \sigmastar$, that is computable and invertible in
polynomial time; this pairing function is extended to encode
$m$-tuples of strings as is standard. For convenience, we
will sometimes write $m$-tuples of strings $x_1, x_2, \ldots , x_m \in
\sigmastar$ explicitly as $x_1 \# x_2 \# \ldots \# x_m$, using a
special separating symbol $\#$ not in~$\Sigma$.
We let $\leq$ denote the standard lexicographic order on~$\sigmastar$.

The definition of Turing machines and their languages, Turing
transducers and the functions they compute, relativized (i.e., oracle)
computations, (relativized) complexity classes, etc.\ is standard in
the literature (see, e.g., the
textbooks~\cite{hop-ull:b:automata,bov-cre:b:complexity,pap:b:complexity}).
We briefly recall the complexity classes most important in this paper.
FP denotes the class of polynomial-time computable functions.  $\fp_1$
is the class of functions computable in polynomial time 
by deterministic transducers with
a unary input alphabet.  FE is the class of functions that can be
computed by deterministic transducers running in time $2^{cn}$ for
some constant~$c$.  Let $\e \equalsdef \bigcup_{c>0}
\mbox{DTIME}[2^{cn}]$ and $\ne \equalsdef \bigcup_{c>0}
\mbox{NTIME}[2^{cn}]$.  An unambiguous Turing machine is a
nondeterministic Turing machine that on each input has at most one
accepting path. UP~\cite{val:j:checking} (respectively, UE) is the
class of all languages accepted by some unambiguous Turing machine
running in polynomial time (respectively, in time $2^{cn}$ for some
constant~$c$).

For any nondeterministic Turing machine $M$ and any input $x \in
\sigmastar$, let $\mbox{\rm acc}_{M}(x)$ denote the number of accepting
paths of~$M(x)$.  A {\em spanP machine\/}~\cite{koe-sch-tor:j:spanp}
is an NP machine 
that has a
special output device on which some output is printed for each
accepting path. For any spanP 
machine $M$ and
any input $x \in \sigmastar$, $\mbox{span}_{M}(x)$ is defined to be
the number of different outputs of $M(x)$ if $M(x)$ has at least one
accepting path, and 0 otherwise.  A {\em tally NP machine\/}
(respectively, a {\em tally spanP machine\/}) is an NP (respectively,
a spanP) machine with a unary input alphabet.

\begin{definition}
\begin{enumerate}
\item {\rm \cite{val:j:permanent,val:j:enumeration}} \quad
$\sharpp \equalsdef \{\mbox{\rm acc}_{M} \mid \mbox{$M$ is
an $\np$ machine} \}$. 

\item {\rm \cite{val:j:enumeration}} \quad
$\sharpp_{1} \equalsdef \{\mbox{\rm acc}_{M} \mid \mbox{$M$ is
a tally $\np$ machine} \}$.

\item {\rm \cite{koe-sch-tor:j:spanp}} \quad
$\spanp \equalsdef \{\mbox{span}_{M} \mid \mbox{$M$ is a $\spanp$
machine} \}$.

\item  $\spanp_{1} \equalsdef \{\mbox{span}_{M} \mid 
\mbox{$M$ is a tally $\spanp$ machine} \}$.

\item 
$\sharpe \equalsdef \{\mbox{\rm acc}_{M} \mid \mbox{$M$ is
an $\ne$ machine} \}$.

\item {\rm \cite{mey-sto:b:reg-exp-needs-exp-space,sto:j:poly}} \quad
The polynomial hierarchy is inductively defined as follows:
$\Sigma_{0}^{p} \stackrel{\mbox{\protect\scriptsize\rm df}}{=} \p$,
$\Sigma_{k}^{p} \stackrel{\mbox{\protect\scriptsize\rm df}}{=}
\np^{\Sigma_{k-1}^{p}}$ 
for $k \geq 1$, and 
$\ph \stackrel{\mbox{\protect\scriptsize\rm df}}{=} 
\bigcup_{i \geq 0} \Sigma_{i}^{p}$.

\item {\rm \cite{gil:j:probabilistic-tms}} \quad
$\pp$ is the class of languages $L$ for which there exist a set 
$A$ in $\p$ and a polynomial $p$ 
such that for all strings $x \in \sigmastar$, 
\[
  x \in L \Lolra \card{\{y \mid |y| = p(|x|) \mbox{ and } \pair{x,y}
  \in A \}} \geq 2^{p(|x|) - 1}.
\]

\item {\rm \cite{gil:j:probabilistic-tms}} \quad
$\bpp$ is the class of languages $L$ for which there exist a set 
$A$ in $\p$ and a polynomial $p$ 
such that for all strings $x \in \sigmastar$, 
\begin{eqnarray*}
x \in L & \Longrightarrow & 
  \card{\{y \mid |y| = p(|x|) \mbox{ and } \pair{x,y} \not\in A \}}
  \leq 2^{p(|x|) - 2}, \mbox{ and } \\ 
x \not\in L & \Longrightarrow & 
  \card{\{y \mid |y| = p(|x|) \mbox{ and } \pair{x,y} \in A
  \}} \leq 2^{p(|x|) - 2}.
\end{eqnarray*}

\item {\rm \cite{cai-hem:j:parity,her:j:mod,bei-gil:j:mod}} \quad
For any fixed $k \geq 2$, ${\rm MOD}_{k}\p$ is the class of languages
$L$ for which there exist a set $A$ in $\p$ and a polynomial $p$ such
that for all strings $x \in \sigmastar$,
\[
  x \in L \Lolra \card{\{y \mid |y| = p(|x|) \mbox{ and } \pair{x,y}
  \in A \}} \not\equiv 0 \mod k.
\]
If $k = 2$, we write $\parityp$ (introduced in {\rm
\cite{pap-zac:c:two-remarks,gol-par:j:ep}}) instead of ${\rm
MOD}_{2}\p$.

\item {\rm \cite{hem-ogi:j:closure,fen-for-kur:j:gap}} \quad
$\spp$ is the class of languages $L$ for which there exist a set 
$A$ in $\p$ and a polynomial $p$ 
such that for all strings $x \in \sigmastar$, 
\begin{eqnarray*}
x \in L & \Longrightarrow & 
  \card{\{y \mid |y| = p(|x|) \mbox{ and } \pair{x,y} \in A \}}
  = 2^{p(|x|) - 1} + 1, \mbox{ and } \\ 
x \not\in L & \Longrightarrow & 
  \card{\{y \mid |y| = p(|x|) \mbox{ and } \pair{x,y} \in A
  \}} = 2^{p(|x|) - 1}.
\end{eqnarray*}

\item {\rm \cite{kar-lip:c:nonuniform}} \quad 
For any language class ${\cal C}$, let ${\cal C}/{\rm poly}$ be the
class of all languages $L$ for which there exist a set $A \in {\cal
C}$, a polynomial $p$, and an advice function~$h : \sigmastar
\rightarrow\, \sigmastar$ such that for each length~$n$, $|h(1^n)| =
p(n)$, and for every $x \in \sigmastar$, $x \in L$ if and only if
$\pair{x, h(1^{|x|})} \in A$.
For any function class ${\cal F}$, let ${\cal F}/{\rm poly}$ be the
class of all functions $g$ for which there exist a function $f \in
{\cal F}$, a polynomial $p$, and an advice function~$h : \sigmastar
\rightarrow\, \sigmastar$ such that for each length~$n$, $|h(1^n)| =
p(n)$, and for every $x \in \sigmastar$, $g(x) = f(\pair{x,
h(1^{|x|})})$.
\end{enumerate}
\end{definition}

We will use the common operator notation at times in order to generalize
function classes such as $\sharpp$ and~$\sharpp_{1}$. 

\begin{definition}
For any language class ${\cal C}$, define
\begin{enumerate}
\item $\# \cdot {\cal C}$ to be the class of functions $f : \sigmastar
\rightarrow\, \nats$ for which there exist a set $A \in {\cal C}$ and
a polynomial $p$ such that for each $x \in \sigmastar$,
\[
f(x) = \card{\{y \mid |y| = p(|x|) \mbox{ and } \pair{x,y} \in A
\}}\mbox{, and }
\]

\item $\#_{1} \cdot {\cal C}$ to be the class of functions $f :
\sigmastar \rightarrow\, \nats$ for which there exist a set $A \in
{\cal C}$ and a polynomial $p$ such that for each $n \in \nats$,
\[
f(1^n) = \card{\{y \mid |y| = p(n) \mbox{ and } \pair{1^n,y} \in A
\}}.
\]
\end{enumerate}
\end{definition}

\begin{definition}
\begin{enumerate}
\item A bijection $\phi : \sigmastar \rightarrow\, \sigmastar$ is a
{\em $\p$-isomorphism\/} if $\phi$ is computable and invertible in
polynomial time.

\item A $\p$-isomorphism $\phi$ is {\em length-preserving\/} if for
all $x \in \sigmastar$, $|\phi(x)| = |x|$.

\item A $\p$-isomorphism $\phi$ mapping set $A \seq \sigmastar$ to set 
$B \seq \sigmastar$ is
{\em order-preserving\/} if for any two strings $x$ and $y$ satisfying 
either $x,y\in A$ or $x,y\not\in A$, if $x \leq y$,
then $\phi(x) \leq \phi(y)$. 

\end{enumerate}
\end{definition}

\begin{definition} 
{\rm \cite{gol-sip:j:compression}}
The {\em ranking function\/} of a language $A \seq \sigmastar$ is the
function $r : \sigmastar \rightarrow\, \nats$ that maps each
$x\in\sigmastar$ to $\vert\,\{ y\leq x \mid y\in A\}\,\vert$.
A language $A$ is {\em rankable\/} if its ranking function is
computable in polynomial time.
\end{definition}

Goldsmith and Homer~\cite{gol-hom:j:scalability} introduced the
property of scalability, a more flexible notion than rankability in
which the rank of some given element within the set is not necessarily
determined with respect to the lexicographic order of~$\sigmastar$,
but rather with respect to {\em any\/} well-ordering of $\sigmastar$
that can be ``scaled'' by a polynomial-time computable and
polynomial-time invertible bijection between $\nats$ and~$\sigmastar$.
Equivalently, the scalable sets are precisely those that are
P-isomorphic to some rankable set.  The definition below is based on this
characterization.

\begin{definition}
{\rm \cite{gol-hom:j:scalability}} A language $A$ is {\em scalable\/} if
it is P-isomorphic to a rankable set.  For any oracle $X$, the
{\em $X$-scalable\/} sets are those that are $\p^X$-isomorphic to some
set rankable in~$\fp^X$.
\end{definition}

\section{Does P Have Easy Census Functions?}
\label{sec:characterizations}

We start with exploring the relationships between the
properties of a set being rankable, being scalable, and having an easy
census function.
Let $A$ be any set (not necessarily in~P). Consider the following conditions:
\begin{desc}
\item[(i)]   $A$ is rankable. 
\item[(ii)]  $A$ has an easy 
        census function.
\item[(iii)]  $A$ is P-isomorphic to some rankable set
        (i.e., $A$ is scalable).
\item[(iv)] $A$ is P-isomorphic to some rankable set via some
length-preserving isomorphism.
\item[(v)]  $A$ is P-isomorphic to some rankable set via
        some order-preserving isomorphism.
\end{desc}

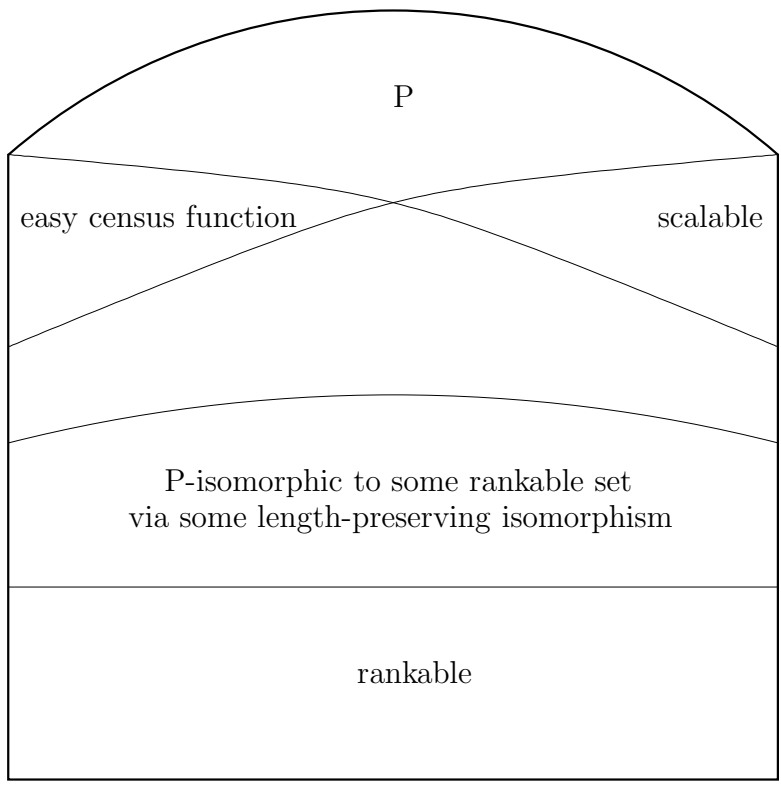
\begin{figure}
\begin{center}
\setlength{\unitlength}{0.00083333in}
\begingroup\makeatletter\ifx\SetFigFont\undefined%
\gdef\SetFigFont#1#2#3#4#5{%
  \reset@font\fontsize{#1}{#2pt}%
  \fontfamily{#3}\fontseries{#4}\fontshape{#5}%
  \selectfont}%
\fi\endgroup%
{\renewcommand{\dashlinestretch}{30}
\begin{picture}(4844,4853)(0,-10)
\thicklines
\put(2422.000,1172.000){\arc{7300.000}{3.9948}{5.4299}}
\thinlines
\put(2422.000,-7328.000){\arc{19500.000}{4.4637}{4.9611}}
\path(22,1222)(4822,1222)
\thicklines
\path(22,3922)(22,22)(4822,22)(4822,3922)
\thinlines
\path(22,3922)(23,3922)(25,3922)
	(30,3921)(36,3921)(46,3920)
	(59,3919)(77,3917)(98,3916)
	(123,3913)(153,3911)(188,3908)
	(227,3904)(271,3901)(319,3896)
	(371,3892)(427,3887)(487,3881)
	(550,3875)(616,3869)(685,3863)
	(756,3856)(828,3849)(902,3842)
	(977,3835)(1052,3827)(1127,3819)
	(1202,3812)(1276,3804)(1349,3795)
	(1422,3787)(1493,3779)(1562,3771)
	(1630,3762)(1697,3754)(1761,3745)
	(1824,3736)(1885,3727)(1945,3718)
	(2002,3708)(2059,3699)(2113,3689)
	(2167,3679)(2219,3668)(2271,3657)
	(2322,3646)(2372,3634)(2422,3622)
	(2470,3610)(2518,3597)(2567,3583)
	(2616,3569)(2667,3554)(2718,3539)
	(2770,3522)(2824,3505)(2878,3486)
	(2935,3467)(2993,3447)(3052,3425)
	(3114,3403)(3176,3380)(3241,3356)
	(3307,3331)(3374,3306)(3443,3279)
	(3512,3252)(3583,3225)(3654,3197)
	(3726,3168)(3798,3139)(3870,3111)
	(3942,3082)(4013,3053)(4082,3025)
	(4151,2998)(4217,2971)(4281,2945)
	(4342,2920)(4400,2896)(4456,2873)
	(4507,2852)(4555,2833)(4598,2815)
	(4638,2798)(4673,2784)(4705,2771)
	(4732,2760)(4754,2750)(4773,2742)
	(4789,2736)(4800,2731)(4809,2727)
	(4815,2725)(4819,2723)(4821,2722)(4822,2722)
\path(4822,3922)(4821,3922)(4819,3922)
	(4814,3921)(4808,3921)(4798,3920)
	(4785,3919)(4767,3917)(4746,3916)
	(4721,3913)(4691,3911)(4656,3908)
	(4617,3904)(4573,3901)(4525,3896)
	(4473,3892)(4417,3887)(4357,3881)
	(4294,3875)(4228,3869)(4159,3863)
	(4088,3856)(4016,3849)(3942,3842)
	(3867,3835)(3792,3827)(3717,3819)
	(3642,3812)(3568,3804)(3495,3795)
	(3422,3787)(3351,3779)(3282,3771)
	(3214,3762)(3147,3754)(3083,3745)
	(3020,3736)(2959,3727)(2899,3718)
	(2842,3708)(2785,3699)(2731,3689)
	(2677,3679)(2625,3668)(2573,3657)
	(2522,3646)(2472,3634)(2422,3622)
	(2374,3610)(2326,3597)(2277,3583)
	(2228,3569)(2177,3554)(2126,3539)
	(2074,3522)(2020,3505)(1966,3486)
	(1909,3467)(1851,3447)(1792,3425)
	(1730,3403)(1668,3380)(1603,3356)
	(1537,3331)(1470,3306)(1401,3279)
	(1332,3252)(1261,3225)(1190,3197)
	(1118,3168)(1046,3139)(974,3111)
	(902,3082)(831,3053)(762,3025)
	(693,2998)(627,2971)(563,2945)
	(502,2920)(444,2896)(388,2873)
	(337,2852)(289,2833)(246,2815)
	(206,2798)(171,2784)(139,2771)
	(112,2760)(90,2750)(71,2742)
	(55,2736)(44,2731)(35,2727)
	(29,2725)(25,2723)(23,2722)(22,2722)
\put(2422,4222){\makebox(0,0)[lb]{\smash{{{\SetFigFont{12}{14.4}{\rmdefault}{\mddefault}{\updefault}P}}}}}
\put(4072,3472){\makebox(0,0)[lb]{\smash{{{\SetFigFont{12}{14.4}{\rmdefault}{\mddefault}{\updefault}scalable}}}}}
\put(2197,622){\makebox(0,0)[lb]{\smash{{{\SetFigFont{12}{14.4}{\rmdefault}{\mddefault}{\updefault}rankable}}}}}
\put(997,1822){\makebox(0,0)[lb]{\smash{{{\SetFigFont{12}{14.4}{\rmdefault}{\mddefault}{\updefault}P-isomorphic to some rankable set}}}}}
\put(772,1597){\makebox(0,0)[lb]{\smash{{{\SetFigFont{12}{14.4}{\rmdefault}{\mddefault}{\updefault}via some length-preserving isomorphism}}}}}
\put(97,3472){\makebox(0,0)[lb]{\smash{{{\SetFigFont{12}{14.4}{\rmdefault}{\mddefault}{\updefault}easy census function}}}}}
\end{picture}
}
\end{center}
\caption{\label{fig:structure} Inclusion structure of the sets in
P satisfying Properties (i) through (iv).}
\end{figure}

It is immediately clear that for any set~$A$, (i) implies each of (ii), (iv),
and~(v), and each of (iv) and (v) implies~(iii). The next proposition
shows that the rankable sets are closed under order-preserving
P-isomorphisms (thus, conditions (i) and (v) in fact are equivalent)
and that the class of sets having an easy census function is closed
under length-preserving P-isomorphisms. The latter fact immediately
gives that (iv) implies~(ii), since each rankable set has an easy
census function.
The inclusion structure of the sets in
P satisfying Properties (i) through (iv) is given in
Figure~\ref{fig:structure}.

\begin{proposition}
\label{prop:order}
\begin{enumerate}
\item \label{prop:order-1} The class of all rankable sets is closed
under order-preserving $\p$-isomorphisms.

\item \label{prop:order-2} The class of sets having an
$\fp$-computable census function is closed under length-preserving
$\p$-isomorphisms.
\end{enumerate}
\end{proposition}

\noindent
{\bf Proof.}  (\ref{prop:order-1}).  Let $A$ be P-isomorphic to a
rankable set $B$ via some order-preserving isomorphism.  Since $B$ is
rankable, $\overline{B}$ is rankable.  Let respectively $r$ and
$\bar{r}$ be the ranking functions for $B$ and~$\overline{B}$. For any
string $x \in \sigmastar$, let $\mbox{lex}(x)$ denote the
lexicographic order of~$x$.  Define the function
\[
  r'(x) \equalsdef \left\{
  \begin{array}{ll}
  r(x) & \mbox{if $x \in A$} \\
  \mbox{lex}(x) - \bar{r}(x) & \mbox{if $x \not\in A$.}
  \end{array}
  \right.
\]
Clearly, $r'$ is computable in polynomial time and $r'$ is the ranking
function for~$A$.

(\ref{prop:order-2}).  Let $A$ be $\p$-isomorphic to a set $B$ with
$\census_B \in \fp$ via some length-preserving
isomorphism $\phi$. Then, $\phi(A^{=n}) = B^{=n}$.  So,
for every $n$, $\census_A = census_B$.  This implies
$\census_B \in \fp$.
\qed

\medskip

So we are left with only the four conditions (i) to~(iv). Since there
are nonrecursive sets with an FP-computable census function, but any
set satisfying one of (i), (iii), or (iv) is in~P, condition
(ii) in general cannot imply any of the other three conditions. On the
other hand, when we restrict our attention to the sets in P having
easy census functions, we can show that (ii) implies (i) if and only
if $\p = \pp$. Thus, even when restricted to P sets, it is unlikely
that (ii) is equivalent to~(i).

\begin{theorem}
All $\p$ sets with an easy census function are rankable if and
only if $\p = \pp$.
\end{theorem}

\noindent
{\bf Proof.}
Hemaspaandra and Rudich show that
$\p = \pp$ (which is equivalent to $\p^{\sharpp} =
\p$) implies that every P set is rankable~\cite{hem-rud:j:ranking}.
Conversely, let $L$ be any set in $\pp$, and let $A$ be a set in P
and $p$ be a polynomial such that
for all~$x \in \sigmastar$,
\[
x \in L \Lolra \card{\{y \mid |y| = p(|x|) \mbox{ and } x\#y \in A
\}} \geq 2^{p(|x|)-1}.
\]
Define
\[
T \equalsdef \{ b\#x\#y \mid x, y \in \sigmastar,\ |y|=p(|x|),\ b
\in \{0,1\}, \mbox{ and } \chi_{A}(x\#y) = b\}.
\]
Clearly, $T \in \p$. Also, the census function of $T$ is easy to
compute: Given $n$ in unary, compute the largest integer $i$ such that
$i+p(i)+3 \leq n$. Then,
\[
\census_{T}(1^{n}) = \left\{
\begin{array}{ll}
2^{i+p(i)} & \mbox{if $i+p(i)+3 = n$} \\
0 & \mbox{if $i+p(i)+3 < n$.}
\end{array}
\right.
\]  
Since $T \in \p$ and $\census_T \in \fp$, by hypothesis $T$ is
rankable.  Let $r$ be the ranking function for~$T$.  Since for each~$x
\in \Sigma^{+}$,
\[
x \in L \Lolra r(0\#x\#1^{p(|x|)}) -
r(1\#\widehat{x}\#1^{p(|\widehat{x}|)}) < 2^{p(|x|)-1},
\]
where $\widehat{x}$ is the lexicographic predecessor of~$x$, and since
the predicate on the right-hand side of the above equivalence can be
decided in polynomial time, it follows that $L \in \p$.~\qed

\begin{corollary}
\label{cor:iff}
All $\p$ sets are rankable if and only if all sets in $\p$ with
an easy census function are rankable.
\end{corollary}

One might ask whether or not all P sets outright have an easy census
function (which, if true, would make Corollary~\ref{cor:iff}
trivial). The following characterization of this question in terms of
unlikely collapses of certain function and language classes suggests
that this probably is not true. Thus, Corollary~\ref{cor:iff} is
nontrivial with the same certainty with which we believe that for instance
not all $\sharpp_{1}$ functions are in~FP\@.\footnote{\protect\singlespacing
It is not
difficult to construct---by standard techniques---an oracle relative
to which $\sharpp_{1} \not\seq \fp$. On the other
hand, we will show in Section~\ref{sec:oracles} that, relative to some
oracle, $\sharpp_{1} \seq \fp$, yet $\sharpp \neq
\fp$ (and thus $\pp \neq \p$).  }

\begin{theorem}
\label{thm:equ1}
The following are equivalent.
\begin{enumerate}
\item \label{equ1-1} Every $\p$ set has an $\fp$-computable census function.
\item \label{equ1-2} $\sharpp_{1} \seq \fp$.
\item \label{equ1-3} $\sharpe = \fe$.
\item \label{equ1-4} $\p^{\sharpp_{1}} = \p$.
\item \label{equ1-5} For every language $L$ accepted by a
logspace-uniform depth~{\rm 2} AND-OR circuit family of bottom fan-in~{\rm 2},
$\census_L$ is in $\fp$.
\end{enumerate}
\end{theorem}

\noindent
{\bf Proof.} To show that (\ref{equ1-1}) implies (\ref{equ1-2}), let
$f$ be any function in $\sharpp_{1}$. Let $M$ be some tally NP machine
with $\mbox{\rm acc}_{M} = f$. Assume that $M$ runs in time~$n^k$, for
some constant~$k$. Define
\[
A \equalsdef \{ x \mid |x| = n^k \mbox{ for some $n$ and $x$
encodes an accepting path of $M(1^n)$} \}.
\]
Clearly, $A$ is in P (note that $n$ can be found in polynomial time,
since computing the $k$th root of some
integer can be done in polynomial time). Now from our hypothesis it
follows that $\census_A$ is in FP, and since $\census_A =
\mbox{\rm acc}_{M}$, we have $f \in \fp$.

Conversely, let $A$ be an arbitrary set in $\p$. Define $M$ to be
the tally NP machine that, on input $1^n$, guesses an $x \in
\{0,1\}^n$, and for each $x$ guessed, accepts along the path for 
$x$ if and only 
if $x \in A$. Then, $\mbox{\rm acc}_{M} = \census_A$.  Since
by hypothesis $\mbox{\rm acc}_{M} \in \fp$, it follows that
$\census_A \in \fp$.

The equivalence of (\ref{equ1-2}) and (\ref{equ1-3}) can be proven
by means of standard translation---this is essentially the function analog
of Book's result that every tally NP set is in $\p$ if and only if 
$\ne = \e$~\cite{boo:j:tally} 
(see~\cite{har:j:upward,har-imm-sew:j:sparse} for the 
extension of this result to sparse sets).

The equivalence of (\ref{equ1-2}) and (\ref{equ1-4}) is 
straightforward.

It is easy to see that (\ref{equ1-2}) implies (\ref{equ1-5}).
In order to prove that (\ref{equ1-5}) implies (\ref{equ1-2}), 
note that computing the number
of satisfying assignments for monotone 2CNF formulas is complete for
$\sharpp$~\cite{val:j:enumeration} under logspace reductions.
Now, given a function $f$ in $\sharpp_1$, there exist logspace
computable functions $R, S, \rho$ such that for all $n$,
$R(1^n)$ is a monotone 2CNF formula with $\rho(1^n)$ variables,
and $f(1^n)$ equals the number of satisfying assignments for
$R(1^n)$ divided by $S(1^n)$.
The reduction $R$ can be modified so that for every $n$,
$\rho(1^{n+1}) > \rho(1^n)$.
Now let $C_m$ be the circuit defined as follows:
(a) if $m = \rho(1^n)$ for some $n$, then $C_m$ is a depth~2
AND-OR circuit that tests whether an assignment, given as the input,
satisfies $R(1^n)$, and (b) if not, $C_m$ is a depth~1 AND circuit
that rejects all inputs.  This circuit family $F = \{ C_m \}$ is
logspace-uniform.  Now let $A$ be the language accepted by $F$.
Then, for every~$n$, $f(1^n) = \census_A(1^{\rho(1^n)})/S(1^n)$.
Thus, (\ref{equ1-5}) implies that $f\in\fp$.~\qed 

\medskip

Theorem~\ref{thm:equ1} can as well be stated for more general classes
than $\sharpp_{1}=\#_{1}\cdot \p$. In particular, this comment
applies to $\#_{1}\cdot {\cal C}$, where for instance ${\cal C} = \np$
or ${\cal C} = \ph$. Noticing that $\spanp_1 = \#_1\cdot \np$ and
focusing on the first two conditions of Theorem~\ref{thm:equ1}, this
observation is  exemplified as follows.


\begin{theorem}
\label{thm:equ2}
\begin{enumerate}
\item Every $\np$ set has an $\fp$-computable census function if and
only if $\spanp_{1} \seq \fp$.
\item Every set in $\ph$ has an $\fp$-computable census function if and
only if $\#_{1} \cdot \ph \seq \fp$.
\end{enumerate}
\end{theorem}

We will show later that the conditions of Theorem~\ref{thm:equ1} in
fact are equivalent to the two conditions stated in either part of
Theorem~\ref{thm:equ2}. 
Next, we give some more evidence that the collapse $\sharpp_1 \seq \fp$ is
unlikely to hold.

\begin{theorem}
\label{thm:impl}
If $\sharpp_{1} \seq \fp$, then the following holds:
\begin{enumerate}
\item For any fixed $k \geq 2$, $\ph \seq {\rm MOD}_{k}\p$, and
\item $\p = \bpp$.
\end{enumerate}
\end{theorem}

\noindent
{\bf Proof.}
For the first part, notice that
Toda and Ogihara~\cite{tod-ogi:j:counting-hard} 
show that for each $k\geq 2$ and any set~$L$, if $L\in\ph$, then
$L \in {\rm MOD}_{k}\p/{\rm poly}$ with an advice 
computable in (the function analog of the language class) 
$\ph^{{\rm MOD}_{k}\p}$.
Also, they show that for every $k\geq 2$, $\ph^{{\rm MOD}_{k}\p} \subseteq
\p^{\sharpp[1]}$, where the $[1]$ in the superscript indicates that 
on every input at most one call to the $\sharpp$ oracle is allowed.
Thus, the advice function for $L$ is in~$\fp_{1}^{\sharpp[1]}$.  
Fix $k \geq 2$ and $L\in\ph$, and take an advice function
$f \in \fp_{1}^{\sharpp[1]}$ that puts $L$ into ${\rm MOD}_{k}\p/{\rm poly}$.
Let $T$ be the polynomial-time oracle transducer with function oracle 
$g \in \sharpp$ that witnesses $f \in \fp_{1}^{\sharpp[1]}$.
W.l.o.g., assume that $T$ makes exactly one oracle call on each
input (by asking a dummy query if necessary). 
Define the $\sharpp_1$ function $g_1$ that, on input~$1^n$, returns
the value~$g(q_n)$, where $q_n$ is the one query string computed by $T$
on input~$1^n$. Thus, $f$ in fact is computable in $\fp_{1}^{\sharpp_{1}[1]}$ 
and so, by our supposition, in polynomial time.  
Since $L$ is in
${\rm MOD}_{k}\p/{\rm poly}$ with polynomial-time computable advice,
it follows that $L \in {\rm MOD}_{k}\p$.
Hence, $\ph \subseteq {\rm MOD}_{k}\p$.

In order to prove the second part, notice that $\bpp$ is in
$\p/{\rm poly}$~\cite{adl:c:two-random} with an advice computable in 
(the function analog of)
PH~\cite{sip:c:randomness,lau:j:bpp}, and that
$\ph \subseteq \p^{\sharpp[1]}$ by Toda's Theorem~\cite{tod:j:pp-ph}.
An argument similar to the above shows that $\p=\bpp$.~\qed

\medskip

Now we show that the conditions of Theorem~\ref{thm:equ1} in fact are
equivalent to the two conditions stated in either part of
Theorem~\ref{thm:equ2}. To this end, we establish the following
theorem, which is interesting in its own right.
Theorem~\ref{thm:stack} is the main technical contribution in this
section.

\begin{theorem}
\label{thm:stack}
$\sharpp_{1}^{\ph} \seq \fp^{\sharpp_{1}^{\sharpp_{1}}}$.
\end{theorem}

\begin{remark}
{\rm 
\begin{enumerate}
\item Note that Toda's result $\ph \seq
\p^{\sharpp[1]}$~\cite{tod:j:pp-ph} immediately gives that
$\sharpp^{\ph} \seq \sharpp^{\sharpp[1]}$ and $\sharpp_{1}^{\ph} \seq
\sharpp_{1}^{\sharpp[1]}$.  Observe that the oracle is a $\sharpp$ function.
In contrast to the inclusion $\sharpp_{1}^{\ph} \seq
\sharpp_{1}^{\sharpp[1]}$,
Theorem~\ref{thm:stack} establishes
containment of $\sharpp_{1}^{\ph}$ in a class in which only
$\sharpp_{1}$ oracles occur. Though our proof also applies the
techniques of~\cite{tod:j:pp-ph,tod-ogi:j:counting-hard}, the result
we obtain seems to be incomparable with the above-mentioned
immediate consequence of
Toda's Theorem. 

\item It is unlikely that Theorem~\ref{thm:stack} can be extended to
$\fp^{\ph}$ or even $\sharpp^{\ph}$ being contained in
$\fp^{\sharpp_{1}^{\sharpp_{1}}}$, since this would imply that
$\fp^{\ph} \seq \fp/{\rm poly}$ and thus, in particular, would
collapse the polynomial hierarchy.  In contrast, the inclusion
$\fp_{1}^{\ph} \seq \fp_{1}/{\rm poly}$ that does follow from (the proof
of) Theorem~\ref{thm:stack} merely implies that all {\em tally\/} sets
in PH have polynomial-size circuits, 
a true statement
that has no unlikely consequences.\footnote{\protect\singlespacing
Indeed, P/poly
is known to contain {\em all\/} tally sets and even the Turing closure
of the sparse sets.
} 

\item The proof of Theorem~\ref{thm:stack} in fact establishes a more
general claim. Since $\parityp^{\ph}/{\rm poly} = \parityp /{\rm
poly}$ \cite{tod-ogi:j:counting-hard}, Theorem~\ref{thm:stack} and its
corollaries can be stated even with PH replaced by $\parityp^{\ph}$
(note that $\parityp^{\ph} = \bpp^{\parityp}$ by Toda's
result~\cite{tod:j:pp-ph}). However, we focus on the PH case, as this
is a more natural and more central class.
\end{enumerate}
}
\end{remark}

\noindent
{\bf Proof of Theorem~\ref{thm:stack}.}  Let $f$ be any function in
$\sharpp_{1}^{\ph}$. Note that 
$\sharpp_{1}^{\ph} = \#_1 \cdot \ph$, since PH is closed under
Turing reductions.  Thus, there exist a set $L \in
\ph$ and a polynomial $p$ such that for each length~$n$, $f(1^n) =
\card{\{y \in \{0,1\}^{p(n)} \mid 1^n\#y \in L\}}$, where for
convenience we assume that $p(n)$ is a power of 2 for each~$n$.  By
Toda and Ogihara's result that $\ph \seq \parityp/{\rm
poly}$~\cite{tod-ogi:j:counting-hard}, there exist a set $A \in
\parityp$, an advice function~$h : \sigmastar \rightarrow\,
\sigmastar$, and a polynomial $q$ such that for each length $m$ and
each $x$ of length~$m$, $|h(1^m)| = q(m)$, and $x \in L$ if and only
if $\pair{x, h(1^m)} \in A$. Let $M$ be a machine witnessing that $A
\in \parityp$, i.e., for every string~$z$, $z \in A$ if and only if
$\mbox{\rm acc}_{M}(z)$ is odd.

Toda~\cite{tod:j:pp-ph} defined inductively the following sequence of
polynomials: For $j \in \nats$, let $s_{0}(j) \equalsdef j$, and for
each $j \in \nats$ and $i > 0$, let 
\[
s_{i}(j) \equalsdef 3(s_{i-1}(j))^4 + 4(s_{i-1}(j))^3.
\] 
One very useful property of this sequence of
polynomials is that for all $i,j \in \nats$, $s_{i}(j) = c \cdot
2^{2^{i}}$ for some $c \in \nats$ if $j$ is even, and $s_{i}(j) = d \cdot
2^{2^{i}} - 1$ for some $d \in \nats$ if $j$ is odd (see~\cite{tod:j:pp-ph}
for the induction proof).

We describe a polynomial-time oracle transducer $T$ that, on input~$1^n$,
invokes its $\sharpp_{1}^{\sharpp_{1}}$ function oracle $g$ and then
prints in binary the number~$f(1^n)$.  Fix the input~$1^n$. First, $T$
transfers the input to the oracle~$g$. Formally, function $g$ is
defined by
\[
g(1^n) \equalsdef \sum_{y \in \{0,1\}^{p(n)}} \left(
s_{\ell_n}(\mbox{\rm acc}_{M}(\pair{1^n\#y, h(1^{n+1+p(n)})})) \right)^2 , 
\]
where $\ell_n \equalsdef \log p(n)$. 

Informally speaking, that
$g$ is in $\sharpp_{1}^{\sharpp_{1}}$ follows from the properties of the
Toda polynomials, from the closure of $\sharpp$ under addition and
multiplication, and from the fact that advice function $h$ is computable
in $\fp_{1}^{\sharpp_{1}[1]}$. 
More formally, to show 
that $g \in \sharpp_{1}^{\sharpp_{1}}$, we describe a tally NP
oracle machine $G$ and a ${\sharpp_{1}}$ oracle $g_1$ for $G$ such that,
for every~$n$, the number of accepting paths of $G$ on input $1^n$
with oracle $g_1$
equals~$g(1^n)$. On input~$1^n$, $G$ first gets the advice string
$a_n = h(1^{n+1+p(n)})$ of length $q(n+1+p(n))$ via
one call to some appropriate ${\sharpp_{1}}$ oracle, say~$g_1$.  
This is possible by the
argument given in the proof of Theorem~\ref{thm:impl}, where $g_1$ is
described. Then, $G$
guesses all strings $y$ of length $p(n)$ and for each $y$ guessed
proceeds as follows. For fixed~$y$, let $j_y$ be a shorthand for 
$\mbox{\rm acc}_{M}(\pair{1^n\#y, a_n})$. Then, $(s_{\ell_n}(j_y))^2$ is a
polynomial of degree $2^{2\ell_n + 1}$, which is polynomial
in~$n$. Also, the coefficients of this polynomial are
deterministically computable in time polynomial in $n$
(see~\cite{tod:j:pp-ph}). Since $\mbox{\rm acc}_{M} \in \sharpp$ and
$\sharpp$ is closed under addition and multiplication, the function
mapping $\pair{1^n\#y, a_n}$ to $(s_{\ell_n}(j_y))^2$ is in~$\sharpp$.
Let $\tilde{G}$ be an NP machine witnessing that 
this function is in~$\sharpp$. Then, $G$ on
input $1^n$ can for each guessed $y$ produce exactly
$(s_{\ell_n}(j_y))^2$ accepting paths by simulating $\tilde{G}$ on
input~$\pair{1^n\#y, a_n}$. Again using the closure of $\sharpp$ under
addition, it follows that $g \in \sharpp_{1}^{\sharpp_{1}}$, as
claimed.

By the above properties of the Toda polynomials, it follows that for
each $y$ of length~$p(n)$, if $j_y$ is even, then $s_{\ell_n}(j_y) =
c \cdot 2^{2^{\ell_n}}$ for some~$c \in \nats$, and if $j_y$ is odd,
then $s_{\ell_n}(j_y) = d \cdot 2^{2^{\ell_n}} - 1$ for some~$d \in
\nats$. Thus, recalling that $2^{\ell_n} = p(n)$, we have
\begin{eqnarray*}
\mbox{$j_y$ is even} & \Longrightarrow & (s_{\ell_n}(j_y))^2 =
(c^{2} \cdot 2^{p(n)-1}) 2^{p(n)+1} \mbox{, and} \\
\mbox{$j_y$ is odd} & \Longrightarrow & (s_{\ell_n}(j_y))^2 =
(d^{2} \cdot 2^{p(n)-1} - d) 2^{p(n)+1} + 1.
\end{eqnarray*}
Defining the integer-valued functions $\widehat{c}(n) \equalsdef c^{2}
\cdot 2^{p(n)-1}$ and $\widehat{d}(n) \equalsdef d^{2} \cdot
2^{p(n)-1} - d$, we obtain
\begin{eqnarray*}
(s_{\ell_n}(j_y))^2 & = &  \left\{
\begin{array}{ll}
\widehat{c}(n) \cdot 2^{p(n)+1}     & \mbox{if $j_y$ is even} \\
\widehat{d}(n) \cdot 2^{p(n)+1} + 1 & \mbox{if $j_y$ is odd.} \\
\end{array}
\right.
\end{eqnarray*}
Thus, since $f(1^n) \leq 2^{p(n)}$ and since $j_y$ is odd if
and only if $1^n\#y \in L$, the rightmost $p(n)+1$ bits of the binary
representation of $g(1^n)$ represent the value of~$f(1^n)$.  Hence,
after the value $g(1^n)$ has been returned by the oracle, $T$ can
output $f(1^n)$ by printing the $p(n)+1$ rightmost bits
of~$g(1^n)$. This completes the proof.~\qed

\medskip

Since $\sharpp_1 \seq \fp$ implies $\fp^{\sharpp_{1}^{\sharpp_{1}}}
\seq \fp$, we have from Theorem~\ref{thm:stack} the following
corollary.

\begin{corollary}
\label{cor:stack}
$\sharpp_1 \seq \fp$ if and only if $\sharpp_{1}^{\ph} \seq \fp$, and in
particular, $\sharpp_1 \seq \fp$ if and only if $\spanp_1 \seq \fp$.
\end{corollary}

Corollary~\ref{cor:stack} together with the equivalences of
Theorems~\ref{thm:equ1} and~\ref{thm:equ2} gives the following.

\begin{corollary}
Every $\p$ set has an easy census function if and only if every
set in $\ph$ has an easy census function.
\end{corollary}

K\"obler et al.~\cite{koe-sch-tor:j:spanp}
proved that $\spanp = \sharpp$ if and only if $\np = \up$. Their proof
also establishes the analogous result for tally sets: 

\begin{lemma}
\label{lem:low}
{\rm (implicit in~\cite{koe-sch-tor:j:spanp})}\quad
$\spanp_{1} = \sharpp_{1}$ if and only if every tally $\np$ set is
in~$\up$.
\end{lemma}

Using Lemma~\ref{lem:low}, we show that $\spanp_1$ and $\sharpp_1$ 
are different classes
unless $\ne = \ue$, or unless every sparse set in $\np$ is low
for~$\spp$.  A set $S$ is said to be ${\cal C}$-{\em low\/} 
for some class ${\cal C}$ if ${\cal C}^S = {\cal C}$ (see, e.g.,
\cite{sch:j:low,ko-sch:j:circuit-low,sch:j:gi,koe-sch-tod-tor:j:few}
for a number of important lowness results).  In particular, it is
known that every sparse NP set is low for
$\p^{\np}$~\cite{ko-sch:j:circuit-low} and for
PP~\cite{koe-sch-tod-tor:j:few}, but it is not known whether all
sparse NP sets are low for~$\spp$. Tor\'{a}n's result that in
some relativized world there exists some sparse NP set that is not contained in
$\parityp$~\cite{tor:thesis:count}, and thus not in~$\spp$, may be
taken as some evidence that not all sparse NP sets are SPP-low.
Since Corollary~\ref{cor:low} relativizes, $\spanp_1 \neq \sharpp_1$ holds
relative to the same oracle.

\begin{corollary}
\label{cor:low}
If $\spanp_1 = \sharpp_1$, then 
\begin{enumerate}
\item $\ne = \ue$ and

\item every sparse $\np$ set is low for~$\spp$.
\end{enumerate}
\end{corollary}

\noindent
{\bf Proof.} The first part follows from a standard upward translation
argument (as mentioned in the proof of Theorem~\ref{thm:equ1}).

For the second part, assume $\spanp_{1} = \sharpp_{1}$, and let $S$ be
any sparse set in~$\np$. Clearly, $S$ polynomial-time truth-table
reduces to some tally NP set~$T$.  By Lemma~\ref{lem:low}, our
assumption implies that $T \in \up$, and thus $T\in \spp$. Since
$\p^{\spp} = \spp$, $S \in \spp$.  The result now follows from the self-lowness
of $\spp$~\cite{fen-for-kur:j:gap}.~\qed

\section{Enumerative Approximation of Census Functions}
\label{sec:enumeration}

Cai and Hemaspaandra~\cite{cai-hem:j:enum} introduced the notion of
enumerative counting as a way of approximating the value of a
$\sharpp$ function deterministically in polynomial time.

\begin{definition}
{\rm \cite{cai-hem:j:enum}} \quad Let $f : \sigmastar \rightarrow\,
\sigmastar$ and $g : \nats \rightarrow\, \nats$ be two functions. A
Turing transducer $E$ is a {\em $g(n)$-enumerator\/} of $f$ if for all
$n \in \nats$ and $x \in \Sigma^{n}$,
\begin{enumerate}
\item $E$ on input $x$ prints a list ${\cal L}_x$ with at most
$g(n)$ elements, and
\item $f(x)$ is a member of list~${\cal L}_x$.
\end{enumerate}

A function $f$ is {\em $g(n)$-enumerable\/} in time $t(n)$ if there
exists a $g(n)$-enumerator of $f$ that runs in time~$t(n)$.

A set is {\em $g(n)$-enumeratively rankable\/} in time $t(n)$ if its
ranking function is $g(n)$-enumerable in time~$t(n)$.
\end{definition}

Recall from the introduction 
Hemaspaandra and Rudich's result that every P set
is $k$-enumeratively rankable for some fixed~$k$ (and indeed, even
${\cal O}(n^{1/2 - \epsilon})$-enumeratively rankable for some
$\epsilon >0$) in polynomial time if and only if 
$\sharpp = \fp$~\cite{hem-rud:j:ranking}. They
conclude that it is no more likely 
that one can enumeratively rank all sets in P
than that one can exactly compute their ranking functions in
polynomial time. We similarly characterize the question of whether the
census function of all P sets is $n^{\alpha}$-enumerable in time
$n^{\beta}$ for fixed constants $\alpha$ and~$\beta$. By the argument
given in the proof of Theorem~\ref{thm:equ1}, this is equivalent to
asking whether every $\sharpp_1$ function is $n^{\alpha}$-enumerable
in time~$n^{\beta}$.  We show that this implies $\sharpp_{1} \subseteq
\fp$, and we thus conclude that it is no more likely that one can 
$n^{\alpha}$-enumerate the census function of every P set in
time~$n^{\beta}$ than that one can precisely compute its census function
in polynomial time.  It would be interesting to know if this result
can be improved to hold for polynomial time instead of time $t$
for some fixed polynomial~$t(n) = n^{\beta}$.

\begin{theorem}
\label{thm:enum}
Let $\alpha, \beta>0$ be constants.
If every $\sharpp_1$ function is $n^{\alpha}$-enumerable in time
$n^{\beta}$, then $\sharpp_{1} \subseteq \fp$.
\end{theorem}

\noindent
{\bf Proof.} 
Cai and Hemaspaandra~\cite{cai-hem:j:approx2}
show that for any fixed~$k$, if $\sharpsat$ (the function mapping any
boolean formula $f$ to the number of satisfying assignments of~$f$)
is $n^k$-enumerable, then
$\sharpp \subseteq \fp$.  In order to prove this, they develop
the following protocol for computing the permanent of an $m\times m$
matrix~$A$, given as parameters (the encoding of) 
a polynomial-time 
transducer $E$ (the enumerator for $\sharpsat$), and a prime number
$p$: Set $A_0 = A$ to the input matrix and repeat the following steps
for $i=1, \ldots, m-1$:
\begin{enumerate}
\item
Construct from $A_{i-1}$ an $(m-i)\times (m-i)$ matrix $B_{i}(X)$,
defined by
\[
B_{i}(X) \equalsdef \sum_{k=1}^{m-i} e_k(X) a_{1k} 
A_{i-1}^{(1,k)},
\]
where $e_k(X)$ is a degree $(m-i)$ polynomial in $X$ such that
$e_k(X) \equiv 1$ if $X=k$ and $0$ otherwise, $a_{1k}$ is the
$(1,k)$ entry of $A_{i-1}$, and $A_{i-1}^{(1,k)}$ is the
$(1,k)$-minor of $A_{i-1}$.  Each matrix is viewed as a matrix over
$\integers/ p\integers$, that is, the matrix entries are 
reduced modulo~$p$.
Then the following conditions hold.
\begin{itemize}
\item
Each entry of $B_i(X)$ is a degree $(m-i)$ polynomial in $X$ with
coefficients in $\{ 0, \ldots, p-1 \}$, so
$\PERM(B_i(X))$ is a degree $(m-i)^2$ polynomial in~$X$.
\item
$\sum_{k=1}^{m-i} \PERM(B_i(k)) = \PERM(A_{i-1})$.
\end{itemize}
\item
Encode $B_i(X)$ into a binary string specifying in binary $p$, $m$,
and the coefficients of $B_i(X)$.  There is some fixed constant $c>0$
such that the encoding length is at most $c(m-i)^3\log p$.
Define $Q_i(X) \equalsdef \PERM(B_i(X))$.  Then, $Q_i$ 
is a polynomial of degree
at most $(m-i)^2$, whose coefficients are each length-bounded by
a fixed polynomial in $p$ and~$m$.  Thus, there is a $\sharpp$
function $G$ that maps $B_i(X)$ to a number from which the coefficients
of $Q_i$ can be decoded in polynomial time.
\item
Use $E$ as an enumerator for $G$ to obtain candidates
$g_1, \ldots, g_t$.  These are all degree $(m-i)^2$ polynomials that
are pairwise distinct.  Since two distinct degree $(m-i)^2$ polynomials
can agree at no more than $(m-i)^2-1$ points, there are fewer than
$t^2(m-i)^2 \leq t^2m^2-1$ points $X$ at which any two candidate
polynomials agree.
Thus, if $p \geq t^2m^2$, then there is an $r\in\{0,\ldots, p-1\}$
such that $g_j(r)\neq g_k(r)$ for all $j\neq k$.  Take the smallest
such $r$ and set $A_i$ to $B_i(r)$ with the entries reduced modulo $p$.
Now, $\PERM(A_i)$ modulo $p$ specifies which $g_j$ is correct, so we
can recover $\PERM(A_{i-1})$ modulo $p$ in polynomial time.
\end{enumerate}
At the end of this loop, $A_m$ is a $1\times 1$ matrix, so its
permanent is easy to compute.  Now working backwards again, we can
recover $\PERM(A)$ modulo $p$.  If we do this for polynomially (in the
encoding length of $A$) many distinct primes, then by the Chinese
Remainder Theorem, we can recover the exact value of $\PERM(A)$.

Valiant~\cite{val:j:permanent} showed that the permanent of matrices 
whose entries are from
the set $\{ -1, 0, 1, 2 \}$ is complete for~$\sharpp$.  
Analogously, we can show that there exists an infinite sequence of
matrices $[ M_1, M_2, \ldots ]$ such that (i)~the mapping
$1^n \rightarrow\, \PERM(M_n)$ is complete for $\sharpp_1$,
(ii)~the mapping $1^n \rightarrow\, M_n$ is polynomial-time
computable, and (iii)~for every $n$, $M_n$ is an $n\times n$
matrix whose entries are from $\{ -1, 0, 1, 2 \}$.  Because
of~(iii), $\PERM(M_n) \leq 2^{2n}$ for all~$n$.  So, by the Chinese
Remainder Theorem, for every~$n$, the exact value of $\PERM(M_n)$ can
be computed from $\PERM(M_n)$ modulo $p$ for $2n$ arbitrary distinct
primes~$p$.  Define polynomials $q$ and $s$ by
$q(n) = \pair{n,n,n,2n}$ and
$s(n) = q(n)^{2\alpha} n^2$.
Define the function $f$ from the tally strings to the set of
natural numbers as follows.
\begin{itemize}
\item
If $m=\pair{H,n,i,j}$ for some $H$,
$i\leq n$ and $j\leq 2n$,
then $f(1^m)$ is $G(B_i(X))$ defined in the above protocol when
we simulate the protocol under the following constraints:
\begin{itemize}
\item
The $j$th smallest prime $> s(n)$ 
is used in place of~$p$.
\item
$M_n$ is used in place of the input matrix~$A_0$.
\item
$H$ is viewed as (the encoding of) a Turing 
transducer and is used in place of the
enumerator~$E$.  Here, for each~$k$ with $1\leq k\leq i-1$, the
input given to $H$ in the $k$th round of the protocol is
$\pair{H,n,k,j}$,
not the matrix $A_k$.  Also, $H$ is
supposed to run in $q(n)^{\beta}$ steps and generates at most
$q(n)^{\alpha}$ candidates in each round.  If $H$ does not halt
in $q(n)^{\beta}$ steps or generates more than $q(n)^{\alpha}$
candidates at any point of the simulation, then the simulation is
immediately aborted and the value $f(1^m)$ is set to~$0$.
\end{itemize}
\item
If $m$ is not of the above form, $f(1^m)$ is $0$.
\end{itemize}
This function $f$ is in $\sharpp_1$.  First, there are only $i\leq m$
rounds to be simulated and each round requires $m^{\alpha}$ steps for
candidate generation and some polynomial (in $n$) number of steps for
other computations.  Second, by the Prime Number Theorem, the $2n$th
smallest prime $> n$ is ${\cal O}(n)$, so finding the 
$j$th smallest prime $> s(n)$ 
requires only a polynomial number of steps.

Now, by our assumption, there is an $m^{\alpha}$-enumerator 
$\widehat{E}$ for $f$ that runs in time $m^{\beta}$.  Since the number
of candidates that $\widehat{E}$ generates is at most 
$m^{\alpha}$ 
and
the dimension of the matrix $M_n$ is~$n$, we have a prime 
$> m^{2\alpha} n^2$.
This
implies that with $\widehat{E}$ as the enumerator, for every $n\geq
\widehat{E}$, every $j, 1\leq j\leq 2n$, and every $i, 1\leq i\leq n$,
we successfully find an $r$ for distinguishing the candidates.  So, with
$\widehat{E}$ as the enumerator, for all $n\geq \widehat{E}$,
$\PERM(M_n)$ is polynomial-time computable.  Hence $\sharpp_1
\subseteq \fp$.~\qed

\section{Oracle Results}
\label{sec:oracles}

In this section, we provide a number of relativized results on the
existence or non-existence of P sets simultaneously satisfying pairs
of conditions chosen among the properties~(i), (ii), and~(iii) from
Section~\ref{sec:characterizations}. For instance,
Theorem~\ref{thm:easy-rank} and its Corollary~\ref{cor:easy-rank}
below exhibit a relativized world in which every P set has an easy
census function (Property~(ii)), yet there exists some set in P that
is not rankable (Property~(i)).

\begin{theorem}
\label{thm:easy-rank}
There exists an oracle $D$ such that $\sharpp_{1}^{D} \seq \fp^{D}
\neq \sharpp^{D}$.
\end{theorem}

From the relativized versions of Theorem~\ref{thm:equ1} and of
Hemaspaandra and Rudich's result in~\cite{hem-rud:j:ranking} 
that every P set is rankable if and only if $\p^{\sharpp} = \p$ (which is
equivalent with $\fp = \sharpp$, and this equivalence itself also
relativizes), we immediately obtain the following corollary.

\begin{corollary}
\label{cor:easy-rank}
There exists an oracle $D$ such that all sets in $\p^{D}$ have a
census function computable in~$\fp^{D}$, yet there exists some set in
$\p^{D}$ that is not rankable by any function in~$\fp^{D}$.
\end{corollary}

\noindent
{\bf Proof of Theorem~\ref{thm:easy-rank}.}  Balc\'{a}zar et
al.~\cite{bal-boo-sch:j:sparse} and Long and
Selman~\cite{lon-sel:j:sparse} proved that the polynomial hierarchy
does not collapse if and only if it does not collapse relative to
every sparse oracle. Since their proof relativizes (i.e., it applies
to the relativized polynomial hierarchy as well), we have the
following claim:

\begin{claim}
\label{c:sparse}
{\rm \cite{bal-boo-sch:j:sparse,lon-sel:j:sparse}} \quad For every
set~$B$, $\ph^{B}$ does not collapse if and only if for every sparse
oracle~$S$, $(\ph^{B})^{S}$ does not collapse.
\end{claim}

Note that $(\ph^{B})^{S} = \ph^{B \oplus S}$, where 
$X \oplus Y \equalsdef \{0x \mid x \in X\} \cup \{1y \mid y \in Y\}$ 
denotes the join of any two sets $X$ and $Y$.
Fix an oracle $A$ such
that $\ph^A$ does not collapse (such oracles were constructed by
Yao~\cite{yao:c:separating}, H{\aa}stad~\cite{has:j:circuits}, and
Ko~\cite{ko:j:exact} who built on the work of Furst et
al.~\cite{fur-sax-sip:j:parity}). Then, by Claim~\ref{c:sparse} above,
for every sparse set~$S$, $\ph^{A \oplus S}$ does not collapse. So, in
particular, $\p^{A \oplus S} \neq \np^{A \oplus S}$ for every
sparse set~$S$.  Since for every oracle~$B$, $\sharpp^B = \fp^B$
implies $\np^B = \p^B$, we have that $\sharpp^{A \oplus S} \neq \fp^{A
\oplus S}$ for every sparse set~$S$.


So it remains to prove that there exists a sparse set $T$ such that
$\sharpp_{1}^{A \oplus T} \seq \fp^{A \oplus T}$. Then, setting $D = A
\oplus T$ completes the proof.

Assume that our pairing function $\pair{\cdot , \cdot , \cdot}$ is
nondecreasing in each parameter, polynomial-time computable and
invertible, and is one-to-one and onto. Let $N_{1}^{(\cdot)},
N_{2}^{(\cdot)}, \ldots$ be a standard enumeration of all tally NP
oracle machines. For each $i \geq 1$, let $p_i$ be the polynomial time
bound of~$N_i^{(\cdot)}$. Then, the function $f^{(\cdot)}$ defined by
\begin{eqnarray*}
f^{(\cdot)}(1^{\pair{i, n, j}}) & \equalsdef &  \left\{
\begin{array}{ll}
\mbox{\rm acc}_{N_{i}^{(\cdot)}}(1^n)  & \mbox{if $p_{i}(n) < j$} \\
0  & \mbox{otherwise} \\
\end{array}
\right.
\end{eqnarray*}
is a canonical function complete for the class
$\sharpp_{1}^{(\cdot)}$.\footnote{\protect\singlespacing 
See~\cite{val:j:enumeration} for 
{\em natural\/} $\sharpp_{1}$-complete functions. 
}
In particular, for every fixed set~$S$, $f^{(A \oplus S)}$ is complete
for $\sharpp_{1}^{A \oplus S}$. 

The oracle set $T$ is defined in such a way that, 
for any given $m = \pair{i, n, j}$ in
unary, some polynomial-time oracle transducer 
can retrieve the value of $f^{(A \oplus T)}(1^m)$ from its
oracle $A \oplus T$ by asking at most $m$ queries. More formally, we
construct $T$ in stages such that for each $m = \pair{i, n, j}$:
\[
1^{k} 0^{m-k} \# b \in T \Lolra \mbox{$1 \leq k \leq |f^{(A \oplus
T)}(1^m)|$ and the $k$th bit of $f^{(A \oplus T)}(1^m)$ is $b$}.
\]
Since by the above definition, $|f^{(A \oplus T)}(1^m)| < m$ and so,
in particular, $N_{i}^{A \oplus T}(1^n)$ cannot query strings of
length $\geq m$, there is no interference between the stages of the
construction of~$T$. It is easy to see that $T$ is a sparse set satisfying
$\sharpp_{1}^{A \oplus T} \seq \fp^{A \oplus T}$.~\qed



\medskip

Now we construct an oracle relative to which there exists some
scalable set in P whose census function is not easy to compute.

\begin{theorem}
There exists an oracle $A$ such that there exists an $A$-scalable set
$B$ whose census function is not in $\fp^{A}$.
\end{theorem}

\noindent
{\bf Proof.} We will construct $A$ and $B$ in such a way that $B$ is
$\p^{A}$-isomorphic to the set $R \equalsdef \{0x \mid x \in
\sigmastar \}$, which is rankable in FP (and thus in
$\fp^{A}$). For each $n \geq 1$, we have $\census_{R}(1^n) = 2^{n-1}$.
So $\census_{R}$ is easy to compute, but we want $B$ to have a hard 
census function. In light of 
Proposition~\ref{prop:order}.\ref{prop:order-2}, we thus need the
isomorphism, $f$, between $B$ and $R$ be non-length-preserving. In
particular, we will define $f$ so as to satisfy $|f(x)| \leq |x| + 1$
and $|f^{-1}(y)| \leq |y|$ for all $x,y \in \sigmastar$. When $f$ is
defined, we let $B$ be the set $f^{-1}(R)$.  To have $f$ and its
inverse computable in $\fp^{A}$, we encode $f$ and $f^{-1}$ into $A
\equalsdef A_{f} \oplus A_{f^{-1}}$ as follows. For all $x \in
\sigmastar$, $i \geq 1$, and $b \in \{0,1\}$, we ensure that
\begin{eqnarray}
\label{equ:code}
\pair{x,i,b} \in A_{f^{\ast}} & \Longleftrightarrow & \mbox{the $i$th
bit of $f^{\ast}(x)$ is~$b$,}
\end{eqnarray}
where $f^{\ast}$ stands for either $f$ or~$f^{-1}$.  At the same time
we diagonalize against $\fp^{A}$ so as to ensure $\census_{B} \not\in
\fp^{A}$.

Let $T_{1}^{(\cdot)}, T_{2}^{(\cdot)}, \ldots$ be a standard 
enumeration of all deterministic
polynomial-time oracle transducers, and let $p_{1}, p_{2}, \ldots$ be
a sequence of strictly increasing polynomials such that $p_{i}$ bounds
the running time of $T_{i}$ (independent of the oracle used).  By
(\ref{equ:code}) above, implicit in the definition of $f$ and $f^{-1}$
is the definition of~$A$, so it suffices to construct the
isomorphism. The construction of $f$ and $f^{-1}$ is in stages. By the
end of stage~$i$, $f$ will have been defined for all strings of length
up to~$r(i)$, where $r$ will be determined below. Initially, we start
with $r(0) = 0$, and we define $f(\epsilon) = \epsilon$. Stage $i > 0$
of the construction is as follows.

\begin{description}
\item[Stage~{\boldmath $i$}:] Choose $n_i$ to be the smallest integer
such that $n_i > r(i-1)$ and $p_i(n_i) < 2^{n_{i}-2}$.  Let $A'$ be
the subset of $A$ that has been decided by now.  We want to define $f$
so that, eventually, $T_{i}^{A}(1^{n_i}) \neq \census_{B}(1^{n_i})$.
Simulate $T_{i}^{A'}$ on input~$1^{n_i}$. Whenever in this simulation
a string of the form $0\pair{x,i,b}$ whose membership in $A$ has not
yet been decided is queried, we add this string to $A'$ and set the
$i$th bit of $f(x)$ to $b$ unless we have already put
$0\pair{x,i,1-b}$ into~$A$ (and thus have set this bit to~$1-b$), or
unless $i>|x|+1$.  The same comment applies to query strings
$1\pair{y,j,b}$ whose membership in $A$ has not been decided yet and
which may fix the $j$th bit of~$f^{-1}(y)$.  If we added the queried
string to~$A'$, we continue the simulation in the ``yes'' state;
otherwise, in the ``no'' state. In this way, the simulation of
$T_{i}^{A'}(1^{n_i})$ may determine $f$ (and $f^{-1}$) on at most
$p_i(n_i) < 2^{n_{i}-2}$ bits of the strings of length~$n_i$. Thus,
for no $m \geq n_i$ is $f^{-1}$ determined on all strings of length
$m$ in $R$ or $\overline{R}$. Once the value $T_{i}^{A'}(1^{n_i})$ is
computed, there is room to decide $f(x)$ and $f^{-1}(y)$ for all
strings $x$ and $y$ of lengths between $r(i-1)$ and $p_i(n_i)$ so that
$f$ is an isomorphism mapping to $\bigcup_{\ell = r(i-1)}^{p_i(n_i)}
R^{= \ell}$ and such that $census_B(1^{n_i}) \neq T_{i}^{A'}(1^{n_i})$, without
changing the output value of $T_{i}^{A'}(1^{n_i})$. Finally, define
$r(i) = p_i(n_i)$.~\qed
\end{description}

\medskip

Next, we provide an oracle relative to which there exists some set in
P that is neither scalable nor has an easy census function.

\begin{theorem}
\label{thm:non-scalable}
There exists an oracle $D$ such that 
$D \in \p^D$ is not $D$-scalable and its census function is not in
$\fp^{D}$.
\end{theorem}

\noindent
{\bf Proof.} 
This is a simple interweaving of two diagonalizations.  The only
question is how to construct a non-scalable set.  
 
It is known from the work of Goldsmith and 
Homer~\cite{gol-hom:j:scalability} that any sparse set is
scalable if and only if it is rankable, and this holds if and only if it is
P-printable.\footnote{\protect\singlespacing
A set is {\em $\p$-printable\/}~\cite{har-yes:j:computation} if there
exists a polynomial-time transducer $T$ such that for each length~$n$,
$T$ on input~$1^n$ prints a list of all elements of the set up to
length~$n$.  
}
$D$ will be sparse, with at most 2 strings at each length.  We assume
that $(T_{i}^{(\cdot)})_{i \geq 1}$ enumerates $\fp^{(\cdot)}$, and that 
$T_{i}^{(\cdot)}$ runs in time~$n^i$.
 
At stage $2i$, we guarantee that $T_{i}^{D}(1^n)$ does not compute the
rank of $1^n$ in~$D$, where $n$ is chosen large enough that $n^i <
2^n$.  For this~$n$, we put $1^n$ into~$D$.  Compute
$T_{i}^{D}(1^n)$, restraining any oracle strings of length $\geq n$ that it
queries.  By our choice of~$n$, this does not decide $D^{=m}$ for any
$m\geq n$, so we can then put in the appropriate number of strings of
length $n$ for the diagonalization.

At stage $2i+1$ we guarantee that $T_{i}^{D}(1^n)$ does not
compute the census function of~$D$, where $n$ is chosen large enough
that $n^i < 2^n$.  Again, compute $T_{i}^{D}(1^n)$, restraining any oracle
strings of length $\geq n$ that it queries.  By our choice of~$n$,
this does not decide $D^{=m}$ for any $m\geq n$, so we can then put
in the appropriate number of strings of length $n$ for the 
diagonalization.~\qed
 
\medskip

Finally, we show that relative to an oracle, there exists some
non-scalable set in P having an easy census function.

\begin{theorem}
\label{thm:non-printable}
There exists an oracle $A$ such that $A \in \p^A$ is not $A$-scalable and its
census function is in $\fp^{A}$.
\end{theorem}

\noindent
{\bf Proof.} We construct the oracle $A$ so that $A$ has one string of
each length.  For those lengths for which nothing else is decided, we
put in~$1^n$.  Otherwise, we do the following.
 
To make the oracle $A$ non-$A$-scalable, we actually make it 
non-$\p^A$-printable.  At stage~$i$, choose an appropriate length~$n$, and
then compute $T_{i}^{A}(1^n)$.  Whenever it queries a string of length
$\geq n$, restrain the string from the oracle.  If it does anything
except print out $A^{\leq n}$, then put in the first unrestrained
string of each length.  If it correctly prints $A$ up to length~$n$,
then choose an $x$ of each relevant length to include that neither is 
restrained nor printed.~\qed


\medskip

We conclude this section with a remark on a technical difficulty in
proving the following statement: ``There exists an oracle $E$ such
that all sets in $\p^{E}$ have a census function computable in
$\fp^{E}$, but $E \in \p^{E}$ is not $E$-scalable.'' Call this
statement~(S). One might hope to prove (S) by exploiting again the
fact that scalability, rankability, and P-printability are equivalent
properties on the sparse sets~\cite{gol-hom:j:scalability}, which was
useful in the proofs of Theorems~\ref{thm:non-scalable}
and~\ref{thm:non-printable}.  Now, replacing in (S) non-scalability by
non-rankability makes (S) the following stronger version of
Theorem~\ref{thm:easy-rank}: ``There exists a {\em sparse\/} set $E$
such that $\sharpp_{1}^{E} \seq \fp^{E} \neq \sharpp^{E}$.'' However,
since the oracle $D = A \oplus T$ constructed in the proof of
Theorem~\ref{thm:easy-rank} inherently is a {\em non\/}sparse set due
to its $A$ part (and it cannot be made sparse unless one could
separate the unrelativized polynomial
hierarchy~\cite{lon-sel:j:sparse,bal-boo-sch:j:sparse}), this approach
does not work to prove~(S). Therefore, to prove~(S), one would need to
construct a nonsparse set $E$ with the desired properties, and we
leave this as an interesting open issue.

\bigskip

{\samepage 
\noindent {\bf Acknowledgments.} \quad We are deeply indebted to Lance
Fortnow, Lane Hemaspaandra, and Gabriel Istrate for interesting
discussions and for helpful comments and suggestions, and we thank
Eric Allender and Lane Hemaspaandra for pointers to the literature.  
}

{\singlespacing

}


\begin{thebibliography}{HRW97b}

\bibitem[Adl78]{adl:c:two-random}
L.~Adleman.
\newblock Two theorems on random polynomial time.
\newblock In {\em Proceedings of the 19th IEEE Symposium on Foundations of
  Computer Science}, pages 75--83, 1978.

\bibitem[All91]{all:j:lim}
E.~Allender.
\newblock Limitations of the upward separation technique.
\newblock {\em Mathematical Systems Theory}, 24(1):53--67, 1991.

\bibitem[AR88]{all-rub:j:print}
E.~Allender and R.~Rubinstein.
\newblock {P}-printable sets.
\newblock {\em SIAM Journal on Computing}, 17(6):1193--1202, 1988.

\bibitem[BBS86]{bal-boo-sch:j:sparse}
J.~Balc\'{a}zar, R.~Book, and U.~Sch\"{o}ning.
\newblock The polynomial-time hierarchy and sparse oracles.
\newblock {\em Journal of the ACM}, 33(3):603--617, 1986.

\bibitem[BC93]{bov-cre:b:complexity}
D.~Bovet and P.~Crescenzi.
\newblock {\em Introduction to the Theory of Complexity}.
\newblock Prentice Hall, 1993.

\bibitem[BG92]{bei-gil:j:mod}
R.~Beigel and J.~Gill.
\newblock Counting classes: {T}hresholds, parity, mods, and fewness.
\newblock {\em Theoretical Computer Science}, 103(1):3--23, 1992.

\bibitem[BH77]{ber-har:j:iso}
L.~Berman and J.~Hartmanis.
\newblock On isomorphisms and density of {{N}{P}} and other complete sets.
\newblock {\em SIAM Journal on Computing}, 6(2):305--322, 1977.

\bibitem[Boo74]{boo:j:tally}
R.~Book.
\newblock Tally languages and complexity classes.
\newblock {\em Information and Control}, 26:186--193, 1974.

\bibitem[CH89]{cai-hem:j:enum}
J.~Cai and L.~Hemachandra.
\newblock Enumerative counting is hard.
\newblock {\em Information and Computation}, 82(1):34--44, 1989.

\bibitem[CH90]{cai-hem:j:parity}
J.~Cai and L.~Hemachandra.
\newblock On the power of parity polynomial time.
\newblock {\em Mathematical Systems Theory}, 23(2):95--106, 1990.

\bibitem[CH91]{cai-hem:j:approx2}
J.~Cai and L.~Hemachandra.
\newblock A note on enumerative counting.
\newblock {\em Information Processing Letters}, 38(4):215--219, 1991.

\bibitem[FFK94]{fen-for-kur:j:gap}
S.~Fenner, L.~Fortnow, and S.~Kurtz.
\newblock Gap-definable counting classes.
\newblock {\em Journal of Computer and System Sciences}, 48(1):116--148, 1994.

\bibitem[FSS84]{fur-sax-sip:j:parity}
M.~Furst, J.~Saxe, and M.~Sipser.
\newblock Parity, circuits, and the polynomial-time hierarchy.
\newblock {\em Mathematical Systems Theory}, 17:13--27, 1984.

\bibitem[GH96]{gol-hom:j:scalability}
J.~Goldsmith and S.~Homer.
\newblock Scalability and the isomorphism problem.
\newblock {\em Information Processing Letters}, 57:137--143, 1996.

\bibitem[Gil77]{gil:j:probabilistic-tms}
J.~Gill.
\newblock Computational complexity of probabilistic {T}uring machines.
\newblock {\em SIAM Journal on Computing}, 6(4):675--695, 1977.

\bibitem[GP86]{gol-par:j:ep}
L.~Goldschlager and I.~Parberry.
\newblock On the construction of parallel computers from various bases of
  boolean functions.
\newblock {\em Theoretical Computer Science}, 43:43--58, 1986.

\bibitem[GS91]{gol-sip:j:compression}
A.~Goldberg and M.~Sipser.
\newblock Compression and ranking.
\newblock {\em SIAM Journal on Computing}, 20(3):524--536, 1991.

\bibitem[Har83]{har:j:upward}
J.~Hartmanis.
\newblock On sparse sets in {NP}$-${P}.
\newblock {\em Information Processing Letters}, 16:55--60, 1983.

\bibitem[H{\aa}s89]{has:j:circuits}
J.~H{\aa}stad.
\newblock Almost optimal lower bounds for small depth circuits.
\newblock In S.~Micali, editor, {\em Randomness and Computation, {\rm volume~5
  of} Advances in Computing Research}, pages 143--170. JAI Press, Greenwich,
  1989.

\bibitem[Hem89]{hem:j:sky}
L.~Hemachandra.
\newblock The strong exponential hierarchy collapses.
\newblock {\em Journal of Computer and System Sciences}, 39(3):299--322, 1989.

\bibitem[Her90]{her:j:mod}
U.~Hertrampf.
\newblock Relations among {MOD}-classes.
\newblock {\em Theoretical Computer Science}, 74(3):325--328, 1990.

\bibitem[HHH]{hem-hem-hem:jtoappear:downward}
E.~Hemaspaandra, L.~Hemaspaandra, and H.~Hempel.
\newblock A downward collapse within the polynomial hierarchy\typeout{MINOR
  PANIC: fill in year, pages, vol, etc}.
\newblock {\em SIAM Journal on Computing}.
\newblock To appear.

\bibitem[HIS85]{har-imm-sew:j:sparse}
J.~Hartmanis, N.~Immerman, and V.~Sewelson.
\newblock Sparse sets in {N}{P}$-${P}: {EXPTIME} versus {NEXPTIME}.
\newblock {\em Information and Control}, 65(2/3):159--181, 1985.

\bibitem[HJ95]{hem-jha:j:defying}
L.~Hemaspaandra and S.~Jha.
\newblock Defying upward and downward separation.
\newblock {\em Information and Computation}, 121:1--13, 1995.

\bibitem[HJRW]{hem-jia-rot-wat:j:join}
L.~Hemaspaandra, Z.~Jiang, J.~Rothe, and O.~Watanabe.
\newblock Boolean operations, joins, and the extended low hierarchy.
\newblock {\em Theoretical Computer Science}.
\newblock To appear.

\bibitem[HR90]{hem-rud:j:ranking}
L.~Hemachandra and S.~Rudich.
\newblock On the complexity of ranking.
\newblock {\em Journal of Computer and System Sciences}, 41(2):251--271, 1990.

\bibitem[HR97]{hem-rot:j:boolean}
L.~Hemaspaandra and J.~Rothe.
\newblock Unambiguous computation: {B}oolean hierarchies and sparse
  {T}uring-complete sets.
\newblock {\em SIAM Journal on Computing}, 26(3):634--653, June 1997.

\bibitem[HRW97a]{hem-rot-wec:j:easy}
L.~Hemaspaandra, J.~Rothe, and G.~Wechsung.
\newblock Easy sets and hard certificate schemes.
\newblock {\em Acta Informatica}, 34(11):859--879, 1997.

\bibitem[HRW97b]{hem-rot-wec:c:easy-one-way-permutations}
L.~Hemaspaandra, J.~Rothe, and G.~Wechsung.
\newblock On sets with easy certificates and the existence of one-way
  permutations.
\newblock In {\em Proceedings of the Third Italian Conference on Algorithms and
  Complexity}, pages 264--275. Springer-Verlag {\it Lecture Notes in Computer
  Science \#1203}, March 1997.

\bibitem[HU79]{hop-ull:b:automata}
J.~Hopcroft and J.~Ullman.
\newblock {\em Introduction to Automata Theory, Languages, and Computation}.
\newblock Addison-Wesley, 1979.

\bibitem[HY84]{har-yes:j:computation}
J.~Hartmanis and Y.~Yesha.
\newblock Computation times of {NP} sets of different densities.
\newblock {\em Theoretical Computer Science}, 34:17--32, 1984.

\bibitem[KL80]{kar-lip:c:nonuniform}
R.~Karp and R.~Lipton.
\newblock Some connections between nonuniform and uniform complexity classes.
\newblock In {\em Proceedings of the 12th ACM Symposium on Theory of
  Computing}, pages 302--309, April 1980.
\newblock An extended version has also appeared as: Turing machines that take
  advice, {\em L'Enseignement Math\'{e}matique}, 2nd series 28, 1982,
  pages~191--209.

\bibitem[Ko89]{ko:j:exact}
K.~Ko.
\newblock Relativized polynomial time hierarchies having exactly $k$ levels.
\newblock {\em SIAM Journal on Computing}, 18(2):392--408, 1989.

\bibitem[KS85]{ko-sch:j:circuit-low}
K.~Ko and U.~Sch\"{o}ning.
\newblock On circuit-size complexity and the low hierarchy in {NP}.
\newblock {\em SIAM Journal on Computing}, 14(1):41--51, 1985.

\bibitem[KST89]{koe-sch-tor:j:spanp}
J.~K{\"{o}}bler, U.~Sch\"{o}ning, and J.~Tor\'{a}n.
\newblock On counting and approximation.
\newblock {\em Acta Informatica}, 26:363--379, 1989.

\bibitem[KSTT92]{koe-sch-tod-tor:j:few}
J.~K{\"{o}}bler, U.~Sch\"{o}ning, S.~Toda, and J.~Tor\'{a}n.
\newblock Turing machines with few accepting computations and low sets for
  {PP}.
\newblock {\em Journal of Computer and System Sciences}, 44(2):272--286, 1992.

\bibitem[Lau83]{lau:j:bpp}
C.~Lautemann.
\newblock {BPP} and the polynomial hierarchy.
\newblock {\em Information Processing Letters}, 14:215--217, 1983.

\bibitem[Lon85]{lon:j:rest}
T.~Long.
\newblock On restricting the size of oracles compared with restricting access
  to oracles.
\newblock {\em SIAM Journal on Computing}, 14(3):585--597, 1985.
\newblock Erratum appears in the same journal, 17(3):628.

\bibitem[LS86]{lon-sel:j:sparse}
T.~Long and A.~Selman.
\newblock Relativizing complexity classes with sparse oracles.
\newblock {\em Journal of the ACM}, 33(3):618--627, 1986.

\bibitem[Mah82]{mah:j:sparse-complete}
S.~Mahaney.
\newblock Sparse complete sets for {NP}: {S}olution of a conjecture of {B}erman
  and {H}artmanis.
\newblock {\em Journal of Computer and System Sciences}, 25(2):130--143, 1982.

\bibitem[MS72]{mey-sto:b:reg-exp-needs-exp-space}
A.~Meyer and L.~Stockmeyer.
\newblock The equivalence problem for regular expressions with squaring
  requires exponential space.
\newblock In {\em Proceedings of the 13th IEEE Symposium on Switching and
  Automata Theory}, pages 125--129, 1972.

\bibitem[OH93]{hem-ogi:j:closure}
M.~Ogiwara and L.~Hemachandra.
\newblock A complexity theory for closure properties.
\newblock {\em Journal of Computer and System Sciences}, 46(3):295--325, 1993.

\bibitem[Pap94]{pap:b:complexity}
C.~Papadimitriou.
\newblock {\em Computational Complexity}.
\newblock Addison-Wesley, 1994.

\bibitem[PZ83]{pap-zac:c:two-remarks}
C.~Papadimitriou and S.~Zachos.
\newblock Two remarks on the power of counting.
\newblock In {\em Proceedings 6th GI Conference on Theoretical Computer
  Science}, pages 269--276. Springer-Verlag {\it Lecture Notes in Computer
  Science \#145}, 1983.

\bibitem[RRW94]{rao-rot-wat:j:upward}
R.~Rao, J.~Rothe, and O.~Watanabe.
\newblock Upward separation for {F}ew{P} and related classes.
\newblock {\em Information Processing Letters}, 52:175--180, 1994.

\bibitem[Sch83]{sch:j:low}
U.~Sch\"{o}ning.
\newblock A low and a high hierarchy within {N}{P}.
\newblock {\em Journal of Computer and System Sciences}, 27:14--28, 1983.

\bibitem[Sch87]{sch:j:gi}
U.~Sch\"{o}ning.
\newblock Graph isomorphism is in the low hierarchy.
\newblock {\em Journal of Computer and System Sciences}, 37:312--323, 1987.

\bibitem[Sip83]{sip:c:randomness}
M.~Sipser.
\newblock A complexity theoretic approach to randomness.
\newblock In {\em Proceedings of the 15th ACM Symposium on Theory of
  Computing}, pages 330--335, 1983.

\bibitem[Sto77]{sto:j:poly}
L.~Stockmeyer.
\newblock The polynomial-time hierarchy.
\newblock {\em Theoretical Computer Science}, 3:1--22, 1977.

\bibitem[TO92]{tod-ogi:j:counting-hard}
S.~Toda and M.~Ogiwara.
\newblock Counting classes are at least as hard as the polynomial-time
  hierarchy.
\newblock {\em SIAM Journal on Computing}, 21(2):316--328, 1992.

\bibitem[Tod91]{tod:j:pp-ph}
S.~Toda.
\newblock {PP} is as hard as the polynomial-time hierarchy.
\newblock {\em SIAM Journal on Computing}, 20(5):865--877, 1991.

\bibitem[Tor88]{tor:thesis:count}
J.~Tor\'{a}n.
\newblock {\em Structural Properties of the Counting Hierarchies}.
\newblock PhD thesis, Universitat Polit\`{e}cnica de Catalunya, Barcelona,
  Spain, 1988.

\bibitem[Val76]{val:j:checking}
L.~Valiant.
\newblock The relative complexity of checking and evaluating.
\newblock {\em Information Processing Letters}, 5:20--23, 1976.

\bibitem[Val79a]{val:j:permanent}
L.~Valiant.
\newblock The complexity of computing the permanent.
\newblock {\em Theoretical Computer Science}, 8:189--201, 1979.

\bibitem[Val79b]{val:j:enumeration}
L.~Valiant.
\newblock The complexity of enumeration and reliability problems.
\newblock {\em SIAM Journal on Computing}, 8(3):410--421, 1979.

\bibitem[Wel93]{wel:b:knots}
D.~Welsh.
\newblock {\em Complexity: Knots, Colourings and Counting}.
\newblock Cambridge University Press, 1993.

\bibitem[Yao85]{yao:c:separating}
A.~Yao.
\newblock Separating the polynomial-time hierarchy by oracles.
\newblock In {\em Proceedings of the 26th IEEE Symposium on Foundations of
  Computer Science}, pages 1--10, 1985.

\end{thebibliography}
\end{document}